\documentstyle{article}

 \normalmarginpar
\twocolumn
 \makeatletter
 \gdef\@proofbox{\relax}
 \long\def\proofbox#1{\gdef\@proofbox{#1}}
 \proofbox{\small{\sl PhysComp96}\\\ifx\UNDEF\@fullpaper
     Extended abstract\else Full paper\fi\\Draft, \@date}

 \gdef\fullpaper{\gdef\@fullpaper{}}

 \def\affil#1{\\{\small#1\par}}
 \gdef\@author{John Doe1\affil{No-Name University, Shipping Dept.}}
 \long\def\author#1{\gdef\@author{#1}}

 \gdef\@abstract{}
 \long\def\abstract#1{\gdef\@abstract{#1}}

\def\@maketitle{\newpage\leavevmode
  \begin{minipage}[t]{0.30\textwidth}
    \hrule height0pt
    \raggedright
    \mbox{}\par
    \@proofbox
  \end{minipage}\relax
  \begin{minipage}[t]{0.70\textwidth}
    \hrule height0pt
    \raggedleft
    \LARGE\@title\par
    \vskip4pt
    \large\@author
  \end{minipage}
  \vskip8pt
  \ifx\@abstract\@empty\else{\vskip.5em\leftskip1.5in\parskip4pt\small\@abstract\par\vskip.5em}\fi
  \rule{\textwidth}{0.4pt}
  \vskip16pt}

\def\@begintheorem#1#2{\sl \trivlist \item[\hskip \labelsep{\bf #1\ #2}]}
\def\@opargbegintheorem#1#2#3{\sl \trivlist
     \item[\hskip \labelsep{\bf #1\ #2\ (#3)}]}

 \newcommand{\sectlabel}[1]{\label{sect.#1}}

  \def\@arabic#1{\number #1} 

\long\def\@makecaption#1#2{
	\vskip\abovecaptionskip
	\sbox\@tempboxa{{\small #1: #2}}%
	\ifdim\wd\@tempboxa>\hsize
	    {\small #1: #2\par}
	\else
	   \global\@minipagefalse
	   \hbox to\hsize{\hfil\box\@tempboxa\hfil}
	\fi
	\vskip \belowcaptionskip}

\def\figstrut#1{\hbox to\linewidth{\vrule height#1\hfill}}

\def\comppad{\thinspace}
\def\comp{\comppad\begingroup \tt \let\do\@makeother \dospecials 
          \@ifstar{\@scomp}{\@comp}}
\def\@scomp#1{\def\@tempa ##1#1{##1\endgroup\comppad}\@tempa}
\def\@comp{\obeyspaces \frenchspacing \@scomp}

\makeatother

\fullpaper

\title{Information Gain vs.\ State Disturbance \\ in Quantum Theory%
\bigskip}
\author{Christopher A. Fuchs\thanks{This work was supported in part by
NSERC.  Present address: Norman Bridge Laboratory of Physics 12-33,
California Institute of Technology, Pasadena, CA 91125. U.~S.~A.}
\affil{D\'{e}partement IRO\\ Universit\'{e} de
Montr\'{e}al\\ C.~P.  6128, Succursale centre-ville\\ Montr\'eal,
Qu\'ebec, Canada H3C 3J7}}
\date{22 September 1996}

\abstract{The engine that powers quantum cryptography is the principle
that there are no physical means for gathering information about the
identity of a quantum system's state (when it is known to be prepared
in one of a set of nonorthogonal states) without disturbing the system
in a statistically detectable way.  This situation is often mistakenly
described as a consequence of the ``Heisenberg uncertainty
principle.''  A more accurate account is that it is a unique feature
of quantum phenomena that rests ultimately on the Hilbert space
structure of the theory {\it along\/} with the fact that time
evolutions for isolated systems are unitary.  In this paper we shall
explore several aspects of the {\it information--disturbance
principle\/} in an attempt to make it firmly quantitative and flesh
out its significance for quantum theory as a whole.}

\begin{document}

\maketitle

\section{Introduction}

Suppose an observer obtains a quantum system secretly prepared in one
of two nonorthogonal pure quantum states.  Quantum theory dictates
that there is no measurement he can use to certify which of the two
states was actually prepared.  This is well known
\cite[and refs]{Fuchs96a}.  A simple, but less recognized, corollary
is that no interaction used for performing such an
information-gathering measurement can leave both states unchanged in
the process \cite{Bennett92a}. If the observer could completely
regenerate the unknown quantum state after measurement, then---by
making further nondisturbing information-gathering measurements on
it---he would eventually be able to infer the state's identity after
all.\footnote{For a nice pedagogical treatment of these
points, see Ref.~\cite{Busch96}.}

This consistency argument is enough to establish a tension between
information gain and disturbance in quantum theory.  What it does not
capture, however, is the extent of the tradeoff between these two
quantities.  In this paper, we shall lay the groundwork for a
quantitative study that goes beyond the qualitative nature of this
tension.  Namely, we will show how to capture in a formal way the idea
that, depending upon the particular measurement interaction, there can
be a tradeoff between the disturbance of the quantum states and the
acquired ability to make inferences about their identity.  We shall
also explore the extent to which the very existence of this tradeoff
can be taken as a fundamental principle of quantum theory---one from
which unitary time evolution itself might possibly be derived.

The plan of the paper is as follows.  In the next Section, we lay out
a general model (inspired by quantum cryptography
\cite{Bennett84,Bennett92b}) upon which to formalize various notions
of the information--disturbance tradeoff.  Section 3 places these
ideas within the historical context, comparing them to the standard
folklore that quantum mechanical measurements cause ``uncontrollable
disturbances'' and do so because of the Heisenberg uncertainty
principle \cite[p.~445]{Peres93b}.
In Section 4, we carry out the calculations necessary to
develop one particularly simple tradeoff relation for pure quantum
states; this example captures much of the essential physics of the
problem.  In Section 5, we say what little can be said presently about
quantitative tradeoff relations for mixed states.  In particular, we
point out how a ``no-cloning'' theorem does not capture the essence of
the problem for mixed states.  In Section 6, we describe a sense in
which Wigner's theorem \cite{Wigner59,Bargman64} allows the standard 
axiom of unitarity for quantum mechanical evolutions to be replaced 
with the ostensibly weaker one that ``eavesdropping on nonorthogonal 
quantum states causes a disturbance.''

\section{The Model}

The model we shall base our considerations on is most easily described
in terms borrowed from quantum cryptography, from whence it takes its
origin.  Alice randomly prepares a quantum system to be in either a
state $\hat\rho_0$ or a state $\hat\rho_1$.  These states, in the most
general setting, are described by density operators on an
$N$-dimensional Hilbert space for some $N$; there is no restriction
that they be pure states, orthogonal, or commuting for that matter.
After the preparation, the quantum system is passed into a ``black
box'' where it may be probed by an eavesdropper Eve in any way allowed
by the laws of quantum mechanics.  That is to say, Eve may allow
the system to interact with an auxiliary system (which we shall call
her {\it probe})---leaving the probe ultimately in one of two states
$\hat\rho^{\scriptscriptstyle {\rm E}}_0$ or
$\hat\rho^{\scriptscriptstyle {\rm E}}_1$---so that after the systems
have decoupled, she may perform quantum mechanical measurements on
the probe itself.  Because the outcome statistics of the measurement
will then be conditioned upon the quantum state that went into the
black box, the measurement may provide Eve with some information about
the quantum state and may even provide her a basis on which to make an
inference as to the state's identity.  After this manhandling by Eve,
the original quantum system is passed out of the black box and back
into the possession of Alice.\footnote{In this respect, the model
differs from the standard one in quantum key distribution.  There,
Eve eventually passes Alice's quantum system on to a third person 
Bob.  Since we are not interested in key distribution per se in this
work, such an extra participant would be superfluous.}

A crucial aspect of this model is that even if Alice knows the exact
manner in which Eve operates---i.e., her probe's initial state and 
the precise interaction she uses---because the system will have become
entangled\footnote{The term ``entanglement'' was first used by
Schr\"odinger \cite{Schrodinger35} soon after the EPR paper
\cite{Einstein35} made its appearance.  For historical interest, and
for the purpose of bringing it to the attention of a wider audience, 
we point out that Einstein had a fairly clear notion of entanglement 
in his mind as early as 9 July 1931. This is evidenced by a letter
from Ehrenfest to Bohr on that date.  See Jammer's article
\cite{Jammer85} for details.} with the probe, she will necessarily
have to resort to a new 
description of the quantum system after it emerges from the black 
box, say by some $\hat\rho_0^{\scriptscriptstyle{\rm A}}$ or
$\hat\rho_1^{\scriptscriptstyle{\rm A}}$.  That is to say, even if
Alice sends a pure state into the black box, a pure state will not
emerge out of it; rather it will be mixed.  This is simply because,
by assumption, Alice does not have access to all the relevant quantum
systems.

This is where the detail of our work begins.  Eve now has the 
potential to gather information about identity of Alice's preparation,
via the alternate states of her probe
$\hat\rho^{\scriptscriptstyle {\rm E}}_0$ or
$\hat\rho^{\scriptscriptstyle {\rm E}}_1$.
Meanwhile the states $\hat\rho_0$ and $\hat\rho_1$ no longer form
valid descriptions of Alice's system because it will have become 
entangled with Eve's ancilla.

The ingredients required to pose a precise {\it tradeoff principle\/}
for information gain and disturbance follow from the setup of the
model.  We shall need:
\begin{itemize}
\item
a concise account of all probes and interactions that Eve may use to
obtain evidence about the identity of the state
\item
a convenient description of the most general kind of quantum
measurement she may then perform on her probe
\item
a measure of the information or inference power provided by any given 
measurement,
\item
a good notion by which to measure the distinguishability of mixed
quantum states and a measure of disturbance based on it, and finally
\item
a ``figure of merit'' by which to compare the disturbance with the
inference.
\end{itemize}

We describe each of these ingredients in some detail in the
subsections below.  Finally in Section 2.5, we explore various ways
of putting them together.  

\subsection{Evolutions}

Since Eve cannot know which of the two quantum states Alice prepared,
the most general manipulation she can perform on her newly acquired
quantum system is to interact it with another system prepared
independently of the particular state coming into the black box.
Thus, if the state of the system entering the black box $\hat\rho_s$,
$s=0,1$, is a density operator on the Hilbert space ${\cal H}_A$, the
initial state of the joint system consisting of it and Eve's probe
will be $\hat\rho_s\otimes\hat\sigma$ for some standard density
operator $\hat\sigma$ on the probe's Hilbert space ${\cal H}_E$.
(There need be no {\it a priori\/} relation between the 
dimensionalities of ${\cal H}_A$ and ${\cal H}_E$.)

Interacting the two systems in the most controlled way that Eve
possibly can, i.e., through some unitary interaction $\hat U$, leads
to the two systems being entangled with only a single quantum state 
on ${\cal H}_A\otimes{\cal H}_E$ between them:
\begin{equation}
\hat\rho^{\scriptscriptstyle {\rm AE}}_s=\hat U(\hat\rho_s\otimes
\hat\sigma)U^\dagger\;.
\end{equation}
After Alice's original system emerges from the black box and is placed
back into her hands, her description of it will be
\begin{equation}
\hat\rho^{\scriptscriptstyle {\rm A}}_s={\rm tr}_{\scriptscriptstyle
{\rm E}}(\hat\rho^{\scriptscriptstyle {\rm AE}}_s)=
{\rm tr}_{\scriptscriptstyle {\rm E}}\Big(
\hat U(\hat\rho_s\otimes\hat\sigma)U^\dagger\Big)\;,
\label{OppieJ}
\end{equation}
where ${\rm tr}_{\scriptscriptstyle {\rm E}}$ represents a partial
trace over Eve's probe.  The system in Eve's possession, will likewise
afterward be described by a new density operator
\begin{equation}
\hat\rho^{\scriptscriptstyle {\rm E}}_s={\rm tr}_{\scriptscriptstyle
{\rm A}}(\hat\rho^{\scriptscriptstyle {\rm AE}}_s)=
{\rm tr}_{\scriptscriptstyle {\rm A}}\Big(
\hat U(\hat\rho_s\otimes\hat\sigma)U^\dagger\Big)\;,
\label{OppieF}
\end{equation}
where ${\rm tr}_{\scriptscriptstyle {\rm A}}$ represents a partial
trace over Alice's system.

The class of evolutions described by Eq.~(\ref{OppieJ}) are called
``nonselective operations'' by Kraus \cite{Kraus83}.  Note that if
$|E_i\rangle$ is a basis in which the operator $\hat\sigma$ is
diagonal, then we may write Eq.~(\ref{OppieJ}) as
\begin{equation}
\hat\rho^{\scriptscriptstyle {\rm A}}_s=\sum_{i,j}
\sqrt{\sigma_j}\langle E_i|\hat U|E_j\rangle\hat\rho_s
\langle E_j|\hat U^\dagger|E_i\rangle\sqrt{\sigma_j}\;,
\label{Sambo}
\end{equation}
where $\sigma_j$ are the eigenvalues of $\hat\sigma$.  Letting the
operators $\hat A_{ij}$ on ${\cal H}_A$ be defined by
\begin{equation}
\hat A_{ij}=\sqrt{\sigma_j}\langle E_i|\hat U|E_j\rangle
\end{equation}
and lumping the indexes, i.e., $\ell=(i,j)$, we can write
Eq.~(\ref{Sambo}) as
\begin{equation}
\hat\rho^{\scriptscriptstyle {\rm A}}_s=\sum_\ell
\hat A_\ell\hat\rho_s\hat A_\ell^\dagger\;.
\label{Louis}
\end{equation}
Note, furthermore, that the operators $\hat A_\ell$ satisfy the
normalization condition
\begin{equation}
\sum_\ell\hat A_\ell^\dagger\hat A_\ell=
\hat I_{\scriptscriptstyle {\rm A}}\;,
\label{Elmo}
\end{equation}
where $\hat I_{\scriptscriptstyle {\rm A}}$ is the identity operator 
on ${\cal H}_A$.  The content of Kraus's representation theorem is 
that {\it any\/} set of operators $\hat A_\ell$ satisfying 
Eq.~(\ref{Elmo}) can be used in Eq.~(\ref{Louis}) to describe a valid
nonselective operation.  (A simple proof of this theorem and a few 
others concerning nonselective operations can be found in 
Ref.~\cite{Schumacher96}.)

The representation of a nonselective operation given by 
Eqs.~(\ref{Louis}) and (\ref{Elmo}) can be quite convenient
because it gives a means of describing the evolution of Alice's system
purely in terms of operators on her own Hilbert space ${\cal H}_A$.  

\subsection{Measurements}

The most general notion of measurement allowed within quantum
mechanics is the POVM (short for {\bf P}ositive {\bf O}p\-erator-{\bf
V}alued {\bf M}easure) \cite{Kraus83,Peres93b}.  These measurements
can always be interpreted in terms of the standard von Neumann type
if one is again willing to make use of an auxiliary system or 
{\it ancilla}.  The idea is simple and much like that considered in 
the last subsection.  Rather than making a von Neumann measurement on 
${\cal H}_E$ directly, there is nothing to stop Eve from first 
interacting her probe with some ancilla and then performing a standard measurement on the newly introduced ancilla.  For instance,
in this way, Eve can perform a valid quantum mechanical measurement
on her probe with a cardinality for the outcome set in excess of the 
dimensionality of ${\cal H}_E$.  Such a type of measurement is 
sometimes necessary for gaining the maximum information about a 
quantum state \cite{Holevo73c,Davies78}.

The nice thing about the formalism of POVMs is that this whole process
can be described formally without ever making any mention of the
extra ancilla.  This can be seen as follows.  Let $\hat\sigma_{\rm c}$
on ${\cal H}_C$ be the initial state of the new ancilla.  After
interacting the probe and the ancilla via some unitary interaction 
$\hat V$, say, the state of the combined system will be
$\hat V(\hat\rho^{\scriptscriptstyle {\rm E}}_s\otimes
\hat\sigma_{\rm c})
\hat V^\dagger$.  If a von Neumann measurement is performed on the
ancilla alone, there is some set of orthogonal projectors
$\hat\Pi_b$ on the ancilla's Hilbert space such that
\begin{equation}
p_s(b)={\rm tr}\Big(\hat V(\hat\rho^{\scriptscriptstyle {\rm E}}_s
\otimes\hat\sigma_{\rm c})\hat V^\dagger(
\hat I_{\scriptscriptstyle {\rm E}}\otimes\hat\Pi_b)\Big)
\end{equation}
is the probability of the various outcomes of that measurement.  Here
$\hat I_{\scriptscriptstyle {\rm E}}$ is the identity operator on
${\cal H}_E$.

Using the cyclic property of the trace function and splitting the
single trace into a piece over the probe and a piece over the ancilla, this becomes
\begin{equation}
p_s(b)={\rm tr}_{\scriptscriptstyle {\rm E}}\Bigg(
\hat\rho^{\scriptscriptstyle {\rm E}}_s\;
{\rm tr}_{\scriptscriptstyle {\rm C}}\!\Big(
(\hat I_{\scriptscriptstyle {\rm E}}\otimes\hat\sigma_{\rm c})
\hat V^\dagger(
\hat I_{\scriptscriptstyle {\rm E}}\otimes\hat\Pi_b)\hat V\Big)
\Bigg)\;.
\end{equation}
Letting
\begin{equation}
\hat E_b = {\rm tr}_{\scriptscriptstyle {\rm C}}\!\Big(
(\hat I_{\scriptscriptstyle {\rm E}}\otimes\hat\sigma_{\rm c})
\hat V^\dagger(\hat I_{\scriptscriptstyle {\rm E}}
\otimes\hat\Pi_b)\hat V\Big)\;,
\end{equation}
we obtain finally that
\begin{equation}
p_s(b)={\rm tr}(\hat\rho_s^{\scriptscriptstyle{\rm E}}\hat E_b)\;.
\end{equation}
As promised, all the operators in this expression refer to the probe's
Hilbert space ${\cal H}_E$.

Note that the operators $\hat E_b$ are all nonnegative definite, i.e.,
\begin{equation}
\langle\psi|\hat E_b|\psi\rangle\ge0
\label{KeyLargo}
\end{equation}
for all $|\psi\rangle$, and satisfy the completeness relation
\begin{equation}
\sum_b\hat E_b=\hat I_{\scriptscriptstyle {\rm E}}\;.
\end{equation}
Any set of operators $\{\hat E_b\}$ satisfying these two properties
is called a POVM.  (This is because these two properties are the
natural generalization to the operator realm of the properties of a
probability measure.)  The POVM formulation of a measurement is
particularly convenient for optimization problems because not only can
POVMs be derived from specific measurement models but, conversely,
any set of operators $\{\hat E_b\}$ satisfying the definition of a
POVM can be identified with a measurement procedure as described
above \cite{Helstrom76}.  This gives an easy algebraic 
characterization of all possible measurements.

\subsection{Information and Distinguishability}

There are several ways by which to quantify the ``information'' that
Eve gathers about the identity of Alice's preparation.  To
say it another way, there are several measures with which to gauge
Eve's performance in learning the identity of $\hat\rho_s$.  The best
measure to use in a tradeoff relation is determined by Eve's
particular needs.  Will Eve be encountering one copy of Alice's
system or many?  If many, then will the systems be prepared the same
each and every time, or rather will they be prepared randomly in one
of the two possibilities at each shot?  If the latter still, must Eve
make an attempt to guess the identity of each state before she 
receives the next, or rather may she wait until very large blocks of
data have been acquired?  All these things and many more must be 
considered before choosing the exact measures for the relation.

For specificity, we shall consider three information and
distinguishability measures \cite{Fuchs96a}.  All these measures, in 
one sense or another, describe how much ``information'' can be gained 
about the identity of a quantum state.  However, only one of the 
measures is truly information theoretic in the sense of Shannon's 
information theory \cite{Shannon48}.  What is common to all three is 
that they each have the power to reveal the physics behind the idea
that information gain and state disturbance go hand in hand for 
quantum physics.

Suppose Eve measures a POVM
$\{\hat E_b\}$ on ${\cal H}_E$.  If Alice prepared $\hat\rho_0$, the
outcomes of Eve's measurement will occur with probabilities
$p_0(b)={\rm tr}(\hat\rho_0^{\scriptscriptstyle{\rm E}}\hat E_b)$; if
Alice prepared $\hat\rho_1$, the outcomes will occur with 
probabilities
$p_1(b)={\rm tr}(\hat\rho_1^{\scriptscriptstyle{\rm E}}\hat E_b)$.
Given this measurement on Eve's part, the extent to which Alice's
preparations can be distinguished is exactly the extent to which
the probability distributions $p_0(b)$ and $p_1(b)$ can be 
distinguished.

A very simple measure of the distinguishability of $p_0(b)$ and 
$p_1(b)$ is the statistical overlap between these distributions
\cite{Wootters81}:
\begin{equation}
B(p_0,p_1)=\sum_b\sqrt{p_0(b)}\sqrt{p_1(b)}\;.
\label{Lee}
\end{equation}
When there is no overlap between the distributions they can certainly
be distinguished completely.  Alternatively, the overlap is unity if
and only if the distributions are identical and cannot be 
distinguished at all.  Such a measure of distinguishability is nice
because of its relative simplicity in expression.  However, this
measure is not completely satisfactory in that it has no direct 
statistical-inferential or information-theoretic meaning.

Another measure of how distinct $p_0(b)$ and $p_1(b)$ are is the
actual Shannon information obtainable about the identity of the
distribution.  For complete generality, let us suppose for the moment
that Eve is expecting the two different quantum states $\hat\rho_0$
and $\hat\rho_1$ with possibly distinct prior probabilities $\pi_0$
and $\pi_1$.\footnote{Up until this point, we have assumed that Eve
has no prior expectation for a bias concerning which state of the two
would be prepared.  That is to say, we have been working
under the assumption that $\pi_0=\pi_1=1/2$.  This remains assumed
even now, despite the further generality of our formulation; the
introduction of an asymmetry is solely for the purpose of comparison
to other forms in the literature.} Then the amount of Shannon (mutual) information 
Eve stands to gain from the measurement $\{\hat E_b\}$ is
\begin{equation}
I(p_0,p_1)=H(p)-\pi_0 H(p_0) - \pi_1 H(p_1)\;,
\label{Harvey}
\end{equation}
where
\begin{equation}
H(q)=-\sum_b q(b)\log q(b)
\end{equation}
is the Shannon entropy of a probability distribution $q(b)$, and,
in particular
\begin{equation}
p(b)\,=\,\pi_0\,p_0(b)\,+\,\pi_1\,p_1(b)
\end{equation}
is the overall prior probability for an outcome $b$.  This measure
is particularly appropriate for gauging Eve's measurement performance
if her purpose is to identify or make an inference about a long string
of quantum states prepared according to the distribution
$\{\pi_0,\pi_1\}$.

Finally, we consider the situation in which Eve bases her performance
on the success of a guess about the identity of one single instance
of Alice's prepared state.  Suppose Eve obtains outcome $b$ in her
measurement.  She can use this knowledge to update her probabilities
or expectations about which state Alice really prepared.  This is
done formally via a use of Bayes' rule \cite{Bernardo94}.  Namely, 
after the measurement, her posterior expectation for the value of $s$
is given by
\begin{equation}
p(s|b)=\frac{\pi_s\, p_s(b)}{p(b)}\;.
\end{equation}
Clearly the best strategy for Eve is to guess the value of $s$ such
that $p(s|b)$ is largest.  Her probability of making an error will
then be the minimum of $p(0|b)$ and $p(1|b)$.  Averaging this over all
possible outcomes gives the expected probability of error upon
making the measurement $\{\hat E_b\}$:
\begin{equation}
P_e(p_0,p_1)=\sum_b\min\Big\{\pi_0\,p_0(b),\,\pi_1\,p_1(b)\Big\}\;.
\label{Oswald}
\end{equation}

The three measures, Eqs.~(\ref{Lee}), (\ref{Harvey}), and 
(\ref{Oswald}), each have the heuristic properties required of one
ingredient in a tradeoff principle:  as we vary across measurements,
the smaller the overlap, the larger the Shannon mutual information,
or the smaller the error probability of a guess, then,
{\it heuristically},
the larger the ``information'' Eve has available about Alice's
preparation.  This is not to say, however, that the measurement 
optimal for any one of these distinguishability measures will be the
same as that for any other; this will generally {\it not\/} be
the case \cite{Fuchs96a}.

For instance, when $\hat\rho_0$ and
$\hat\rho_1$ are both invertible, the measurement optimal for
minimizing the statistical overlap \cite{Fuchs95b}, Eq.~(\ref{Lee}),
is the von Neumann measurement corresponding to the rather complicated
Hermitian operator\footnote{The square root of a nonnegative definite
operator signifies an operator diagonal in the same basis as the said
operator but whose eigenvalues are the positive square roots of the
previous.}
\begin{equation}
\left.\hat\rho_1^{\scriptscriptstyle{\rm E}}
\right.^{\scriptscriptstyle{-}\!\frac{1}{2}}
\sqrt{\left.\hat\rho_1^{\scriptscriptstyle{\rm E}}
\right.^{\!\frac{1}{2}}
\hat\rho_0^{\scriptscriptstyle{\rm E}}
\left.\hat\rho_1^{\scriptscriptstyle{\rm E}}
\right.^{\!\frac{1}{2}}\,}\!
\left.\hat\rho_1^{\scriptscriptstyle{\rm E}}\right.
^{\scriptscriptstyle{-}\!\frac{1}{2}}\;.
\end{equation}

The measurement optimal for maximizing the mutual information, 
Eq.~(\ref{Harvey}), generally has no analytic expression, as can
already be seen at the level of density operators on two dimensional
Hilbert spaces \cite{Fuchs94,Fuchs96a}.  On top of that, more
generally, the number of outcomes in an optimal measurement
is not even known; the strongest result yet proven only bounds that
number to be less than the square of the Hilbert space dimension
\cite{Davies78}.

In contrast, the measurement for optimizing the 
probability of error \cite{Helstrom76}, Eq.~(\ref{Oswald}), is again
a von Neumann measurement.  In fact, this time the optimal measurement
corresponds to a fairly simple Hermitian operator:
\begin{equation}
\hat\Gamma=\pi_1\hat\rho_1^{\scriptscriptstyle{\rm E}}
-\pi_0\hat\rho_0^{\scriptscriptstyle{\rm E}}\;.
\label{BowTie}
\end{equation}
Generally there are no values of $\pi_0$ and $\pi_1$ such that all
three ``optimal'' measurements coincide.

To reflect the fact that we are interested only in Eve's ultimate
performance as an ingredient for an information--disturbance 
principle, we focus hereafter on the optimized versions of the 
three distinguishability measures.  This forces all measurement 
dependence on Eve's part out of the picture.

The minimal statistical overlap after varying over all measurements
turns out to be \cite{Fuchs95b,Fuchs96a}:
\begin{equation}
B(\hat\rho_0^{\scriptscriptstyle{\rm E}},
\hat\rho_1^{\scriptscriptstyle{\rm E}})\,=
{\rm tr}\,\sqrt{\left.\hat\rho_1^{\scriptscriptstyle{\rm E}}
\right.^{\!\frac{1}{2}}
\hat\rho_0^{\scriptscriptstyle{\rm E}}
\left.\hat\rho_1^{\scriptscriptstyle{\rm E}}
\right.^{\!\frac{1}{2}}\;}\,\rule{0mm}{5.5mm}\;.
\label{Hannibal}
\end{equation}
This is quantity is related to the Uhlmann {\it fidelity\/} function
$F(\hat\rho_0^{\scriptscriptstyle{\rm E}},
\hat\rho_1^{\scriptscriptstyle{\rm E}})$ for quantum states
\cite{Uhlmann76,Jozsa94a} through\footnote{A distinction is drawn
between fidelity and minimal statistical overlap because, for
different applications, one form can be more convenient than the
other.  As far as interpretation is concerned, the two quantities
appear to be on an equal footing.}
\begin{equation}
F(\hat\rho_0^{\scriptscriptstyle{\rm E}},
\hat\rho_1^{\scriptscriptstyle{\rm E}})=
\Big(B(\hat\rho_0^{\scriptscriptstyle{\rm E}},
\hat\rho_1^{\scriptscriptstyle{\rm E}})\Big)^{\!2}\;.
\end{equation}

The fidelity has several interesting properties despite its
somewhat loose connection to statistical tests.  For instance,
though it is necessarily bounded between 0 and 1, it equals unity if
and only if the two quantum states in its arguments are
identical.  Also, despite first appearances, it is symmetric in its
arguments.  In the case that
$\hat\rho_1^{\scriptscriptstyle{\rm E}}=|
\psi_1^{\scriptscriptstyle{\rm E}}\rangle\langle
\psi_1^{\scriptscriptstyle{\rm E}}|$
is a pure state, Eq.~(\ref{Hannibal}) reduces to
\begin{equation}
F(\hat\rho_0^{\scriptscriptstyle{\rm E}},
\hat\rho_1^{\scriptscriptstyle{\rm E}})\,=\,
\langle\psi_1^{\scriptscriptstyle{\rm E}}|
\hat\rho_0^{\scriptscriptstyle{\rm E}}
|\psi_1^{\scriptscriptstyle{\rm E}}\rangle\;.
\end{equation}

The optimal mutual information, or {\it accessible information},
$I(\hat\rho_0^{\scriptscriptstyle{\rm E}},
\hat\rho_1^{\scriptscriptstyle{\rm E}})$, again, generally has no
analytic expression.  However, there is a growing literature of 
successively better bounds on this quantity---see references in 
Ref.~\cite{Fuchs96a}.  One particularly interesting bound (because
of its simplicity in form) due to Holevo \cite{Holevo73d,Fuchs94} is
\begin{equation}
I(\hat\rho_0^{\scriptscriptstyle{\rm E}},
\hat\rho_1^{\scriptscriptstyle{\rm E}})
\le S(\hat\rho^{\scriptscriptstyle{\rm E}})-
\pi_0 S(\hat\rho_0^{\scriptscriptstyle{\rm E}})-
\pi_1 S(\hat\rho_1^{\scriptscriptstyle{\rm E}})\;,
\label{DuoTang}
\end{equation}
where
\begin{equation}
\hat\rho^{\scriptscriptstyle{\rm E}}=
\pi_0\hat\rho_0^{\scriptscriptstyle{\rm E}}+
\pi_1\hat\rho_1^{\scriptscriptstyle{\rm E}}
\end{equation}
and 
\begin{equation}
S(\hat\sigma)=-{\rm tr}(\hat\sigma\log\hat\sigma)
\end{equation}
is the von Neumann entropy of a density operator $\hat\sigma$.  It
should be pointed out, however, that this bound is often not very
tight; the satisfaction of equality in Eq.~(\ref{DuoTang}) comes
about if and only if $\hat\rho_0^{\scriptscriptstyle{\rm E}}$ and
$\hat\rho_1^{\scriptscriptstyle{\rm E}}$ commute.

Finally, as might have been expected, the optimal error probability
turns out to be the simplest distinguishability measure of the lot:
\begin{equation}
P_e(\hat\rho_0^{\scriptscriptstyle{\rm E}},
\hat\rho_1^{\scriptscriptstyle{\rm E}})=\frac{1}{2}-
\frac{1}{2}{\rm tr}\Bigl|\,
\pi_1\hat\rho^{\scriptscriptstyle {\rm E}}_1-
\pi_0\hat\rho^{\scriptscriptstyle {\rm E}}_0\,\Bigr|\;,
\label{LimaPeru}
\end{equation}
where $|\cdot|$ signifies an operator diagonal in the same basis as
its argument, but with eigenvalues that are the absolute values of
those of the argument.  (Because this particular distinguishability
measure is so important for the considerations here, we give a new
derivation of Eq.~(\ref{LimaPeru}) in the Appendix.)

Of the three measures of Eve's performance discussed here, perhaps
the accessible information is the most relevant for eavesdropping
in quantum key distribution.  However, even this has its shortfalls.
Presently it is not known whether there is any single numeric summary
of Eve's posterior probabilities for the identity of
$\hat\rho_s$---such as the mutual information---that can
be used for privacy amplification \cite{Bennett95Q} against all 
possible eavesdropping strategies \cite{Mayers95T}.

\subsection{Disturbance Measures}

In order to gauge the amount of disturbance Eve inflicts upon Alice's
system as she gains information about it, we must find a way compare
the state of Alice's system before it enters the black box to the
state of the system after it exits.
That is to say, we must find a way to compare both $\hat\rho_0$ to 
$\hat\rho_0^{\scriptscriptstyle{\rm A}}$ and $\hat\rho_1$ to
$\hat\rho_1^{\scriptscriptstyle{\rm A}}$.  Of course, no
{\it single\/} simple numerical measure or ``distance'' will be
adequate to all tasks:  the quantum states are operators on
${\cal H}_A$.  In matrix representation, each state will have up to
$N^2-1$ real parameters (corresponding to the real and imaginary parts
of its entries).  Just naively, one would expect that it should take 
at least that many numbers to capture all aspects of Eve's 
disturbance.

Nevertheless, one can still go a long way toward quantifying
disturbance with simple numerical measures.  In particular, to get
started, one need only notice that there are at least as many 
disturbance measures as there are distinguishability measures.  Each
of these can be used as a notion of ``distance'' between the initial 
and final states.  Moreover, these are especially nice in that they
have direct experimental significance, each being intrinsically
tied to some statistical or information theoretic test.

As a simple example of how a distinguishability measure can be used
to make a disturbance measure, consider the following quantity:
\begin{equation}
D(\hat U)=
1-\frac{1}{2}F(\hat\rho_0,\hat\rho_0^{\scriptscriptstyle{\rm A}})
-\frac{1}{2}F(\hat\rho_1,\hat\rho_1^{\scriptscriptstyle{\rm A}})\;.
\label{dingus}
\end{equation}
This is just one minus the average fidelity between input and
output.  Notice that $D(\hat U)$ has a functional dependence on the
initial states $\hat\rho_0$ and $\hat\rho_1$, the dimensionality of
Eve's probe's Hilbert space ${\cal H}_E$, the probe's initial state
$\hat\sigma$, and the particular unitary interaction $\hat U$ that
Eve uses for her eavesdropping.  For brevity and convenience, we have
made only the functional dependence on $\hat U$ explicit in the
labeling of this disturbance measure.

Equation~(\ref{dingus}) fulfills---at the very least---the minimal
requirement of an intuitive disturbance measure: it has the virtue
that it vanishes if and only if 
$\hat\rho_0^{\scriptscriptstyle{\rm A}}=\hat\rho_0$ {\it and\/} 
$\hat\rho_1^{\scriptscriptstyle{\rm A}}=\hat\rho_1$.  In addition to
that, however, it equals unity if and only if the outputs are
orthogonal to the inputs---that is, that Eve's eavesdropping has
probability one of being caught.  Finally, because it can be
interpreted as related to the statistical overlap of two probability 
distributions, we can be sure that it is monotonic with our intuition
at all points between.

Eq.~(\ref{dingus}) is perfectly reasonable as far our purposes are
concerned.  Nevertheless, one should not lose sight of the fact that 
$D(\hat U)$ is quite arbitrary as a 
disturbance measure.  This remains true even if one insists that a
disturbance measure be built from the fidelity function alone and
not any other distinguishability measure.  For instance, one can
still ask, ``Why take an average of the two fidelities?  Why not a
geometrical mean?  Why not relate the disturbance to the minimum of 
the two fidelities?''   The questions can go on and on.  The only 
means for relieving the arbitrariness in choosing a disturbance 
measure is to have a specific physical/experimental application in 
mind.  This is just as it was for the discussion of Eve's possible 
information-gain measures.

Let us give one more example along these lines to help firm the idea
that the number of variations on the theme is quite large.  Suppose
Alice expects the eavesdropper to be active at random and only 50\%
of the time.  Whenever active, Eve will use the same strategy
Eq.~(\ref{Louis}), which Alice knows in advance.  A natural notion of
disturbance for this particular scenario is the (best possible)
probability that Alice can catch Eve out when she is active.  When
the probability of catching Eve is large, then the disturbance can be
assumed large; when the probability of catching Eve is no better than
chance, then the disturbance is small.  Thus, referring to
Eq.~(\ref{LimaPeru}), the disturbance of input state $s$ is
quantified by
\begin{equation}
P_s=\frac{1}{2}+\frac{1}{4}{\rm tr}\Big|
\hat\rho_s-\hat\rho_s^{\scriptscriptstyle{\rm A}}\Big|\;.
\end{equation}
On average then, the disturbance viewed in this way is
\begin{eqnarray}
D_{\rm G}(\hat U)
&=&
\frac{1}{2}P_0\,+\,\frac{1}{2}P_1
\nonumber\\
&=&
\frac{1}{2}\,+\,\frac{1}{8}{\rm tr}\Big|
\hat\rho_0-\hat\rho_0^{\scriptscriptstyle{\rm A}}\Big|
\,+\,\frac{1}{8}{\rm tr}\Big|
\hat\rho_1-\hat\rho_1^{\scriptscriptstyle{\rm A}}\Big|\,.
\rule{0mm}{6mm}
\nonumber\\
\end{eqnarray}
The same sort of game, of course, can also be played with the maximal
amount of information that Alice can gain about Eve's activity, giving
rise to still another disturbance measure.

Finally we note that, above and beyond these examples, there are
still other ways of using the distinguishability measures of the last
subsection for the purpose of gauging Eve's disturbance.  For
instance, they may be used not only to compare the input and output 
states of Alice's system but also for checking the extent to which Eve
has broken the coherence of that system with yet other systems.  This 
is an idea due to Schumacher \cite{Schumacher96}.  It is especially 
relevant to the model considered in this paper because Alice recovers 
her quantum system after Eve has interacted with it; Alice's system
is not passed on to a third person, such as Bob, as is the case in 
quantum key distribution scenarios.  Therefore Alice is in the 
position to check whether coherences have indeed been broken.  By the 
same token, this measure may be less relevant to quantum cryptography.

Let us spell out Schumacher's idea in some detail.  Suppose Alice
prepares her two states $\hat\rho_0$ and $\hat\rho_1$ by preparing
a composite system in the pure states
$|\psi_0^{\scriptscriptstyle {\rm AB}}\rangle$ and
$|\psi_1^{\scriptscriptstyle {\rm AB}}\rangle$ on the tensor product
Hilbert space ${\cal H}_A\otimes{\cal H}_B$.  The intention is to give
to Eve only that part of the system described on the Hilbert space
${\cal H}_A$.  Therefore, for $s=0,1$, the pure states must
satisfy
\begin{equation}
\hat\rho_s={\rm tr}_{\scriptscriptstyle {\rm B}}\Big(
|\psi_s^{\scriptscriptstyle {\rm AB}}\rangle
\langle\psi_s^{\scriptscriptstyle {\rm AB}}|\Big)\;.
\end{equation}
If the upshot of Eve's interaction with ${\cal H}_A$ is given by
Eq.~(\ref{Louis}), then the overall state of the composite system
after the single component emerges from the black box will be
\begin{equation}
\hat\rho^{\scriptscriptstyle {\rm AB}}_s=\sum_\ell
(\hat A_\ell\otimes\hat I_{\scriptscriptstyle {\rm B}})
|\psi_s^{\scriptscriptstyle {\rm AB}}\rangle
\langle\psi_s^{\scriptscriptstyle {\rm AB}}|
(\hat A_\ell\otimes\hat I_{\scriptscriptstyle {\rm B}})^\dagger\;,
\end{equation}
where $\hat I_{\scriptscriptstyle {\rm B}}$ is the identity operator
on ${\cal H}_B$.  The main point is that we need not be content 
comparing only $\hat\rho_s$ to
$\hat\rho_s^{\scriptscriptstyle{\rm A}}$ because in this scenario we
also have the option of comparing 
$|\psi_s^{\scriptscriptstyle {\rm AB}}\rangle$ to
$\hat\rho^{\scriptscriptstyle {\rm AB}}_s$.

One might think that such a procedure would open a can of worms with
respect to the original statement of our problem.  For instance, by
introducing the {\it purification\/}
$|\psi_s^{\scriptscriptstyle {\rm AB}}\rangle$ of $\hat\rho_s$, one
might think we are introducing much more freedom in the formulation of
the problem than there was previously.  There are an infinite number
of purifications of the given density operators with all manner of
relations between themselves.  One would think that surely, whatever
disturbance measure we build with these purifications, it will inherit
that extraneous freedom.

Surprisingly, that turns out
{\it not\/} to be the case---at least for some measures of 
distinguishability.  To take as an example, one can speak of the
fidelity between input and output in this scenario, a quantity
Schumacher calls the {\it entanglement fidelity\/}:
\begin{equation}
F_s\equiv F\Big(|\psi_s^{\scriptscriptstyle {\rm AB}}\rangle
\langle\psi_s^{\scriptscriptstyle {\rm AB}}|\,,\,
\hat\rho^{\scriptscriptstyle {\rm AB}}_s\Big)\;.
\end{equation}
This quantity turns out to be completely independent of the particular
purification chosen for $\hat\rho_s$ \cite{Schumacher96}.  Hence it
can be expressed purely in terms of $\hat\rho_s$ and the particular
operation Eve subjected $\hat\rho_s$.  In Ref.~\cite{Schumacher96},
it is shown that
\begin{equation}
F_s=\sum_\ell \Big|{\rm tr}(\hat\rho_s\hat A_\ell)\Big|^2\;.
\end{equation}

As might be expected, the entanglement fidelity can be a more exacting
measure of change between input and output.  For instance, it can be
proven that \cite{Jozsa94a,Barnum96a}
\begin{equation}
F_s\le F(\hat\rho_s,\hat\rho_s^{\scriptscriptstyle{\rm A}})\;.
\end{equation}
Thus, if Eq.~(\ref{dingus}) is considered an adequate disturbance
measure for a particular application of our model, then so too
should the more stringent measure
\begin{eqnarray}
D_{\rm S}(\hat U)
&=&
1\,-\,\frac{1}{2}F_0\,-\,\frac{1}{2}F_1
\nonumber\\
&=&
1\,-\,\frac{1}{2}\sum_\ell
\Big|{\rm tr}(\hat\rho_0\hat A_\ell)\Big|^2\,-\,
\frac{1}{2}\sum_\ell \Big|{\rm tr}(\hat\rho_1\hat A_\ell)\Big|^2\;.
\rule{0mm}{6mm}
\nonumber\\
\label{dingus2}
\end{eqnarray}

This is just a small sampling of the numerous things that can be
imagined for gauging disturbance in a quantifiable way.  Nevertheless,
the options are already great enough to start thinking seriously of
several ways of viewing an information--disturbance tradeoff
principle.

\subsection{Tradeoff}

Finally, we are in a position of taking our two building blocks,
{\it information measures\/} and {\it disturbance measures}, and 
combining them to arrive at a formal tradeoff principle.  The first
thing to recall is that once the two quantum states $\hat\rho_0$ and 
$\hat\rho_1$ are fixed, the only free variables in {\it any\/} of the
measures are the dimensionality of ${\cal H}_E$, the probe's initial
state $\hat\sigma$, and the unitary interaction $\hat U$.
To expedite our discussion, let us consider a generic information
measure ${\cal I}(\hat U)$ and a generic disturbance measure
${\cal D}(\hat U)$.  The discussion to follow remains true for any
two such measures.  It should be noted that only the
$\hat U$-dependence of ${\cal I}$ and ${\cal D}$ has been made 
explicit.  (The reason for this will be made clear below.)

There are at least two straightforward ways to combine  
${\cal I}(\hat U)$ and ${\cal D}(\hat U)$ into various
``tradeoff relations.''  The simplest idea is rather analogous to the
construction of the Heisenberg uncertainty relations in that it is
based on the idea that there must be some standard ``figure of merit''
for comparing information and disturbance.  In the Heisenberg
relations, the figure of merit for the ``tradeoff'' of $\Delta x$
and $\Delta p$ is a multiplicative one:
\begin{equation}
\Delta x\Delta p\ge\frac{\hbar}{2}\;.
\end{equation}
In our case, we may well imagine a standard or convenient function
${\cal F}:{\rm I\kern-.2emR}\times{\rm I\kern-.2emR}\rightarrow
{\rm I\kern-.2emR}$
with which to compare the two
quantities of our concern.  If it is an interesting function, then
it will be bounded in at least one direction---say, from below---and
we will be at liberty to define
\begin{equation}
{\cal B}(\hat\rho_0,\hat\rho_1)
=\min\,{\cal F}\Big({\cal I}(\hat U),{\cal D}(\hat U)\Big)\;,
\label{Hyacinth}
\end{equation}
where the minimum is taken over all the free variables of ${\cal I}$
and ${\cal D}$.  Notice that what remains is a function purely of the
two input states $\hat\rho_0$ and $\hat\rho_1$.
Eq.~(\ref{Hyacinth}) defines an {\it information--disturbance
tradeoff relation\/} for an arbitrary probe, probe initial state, and
interaction on Eve's part by
\begin{equation}
{\cal F}\Big({\cal I}(\hat U),{\cal D}(\hat U)\Big)\ge
{\cal B}(\hat\rho_0,\hat\rho_1)\;.
\label{Pearl}
\end{equation}

The right hand side of Eq.~(\ref{Pearl}) plays the role of the 
$\hbar/2$ in the Heisenberg relations.  As already noted, the lower 
bound in this new principle depends on the two quantum states under
consideration.  This is completely reasonable:  one expects a
necessary information--disturbance tradeoff only when the two states 
being compared cannot be interpreted as ``classical'' states.  For 
instance, when $\hat\rho_0$ and $\hat\rho_1$ are pure and orthogonal,
all the information about the state's identity can be recovered
without incurring any disturbance whatsoever.

The second way to think about a tradeoff relation is to ask, ``For
a fixed disturbance, what is the maximal information that Eve can
recover about the state's identity?''  This defines a
constrained-variation problem:  for fixed
\begin{equation}
{\cal D}(\hat U)=d\;,
\end{equation}
where $d$ is some constant, what is
\begin{equation}
{\cal I}_{\rm opt}=\max\,{\cal I}(\hat U)\;?
\end{equation}
The optimization here again is taken over all the free variables of 
${\cal I}$, namely over all probes, over all initial states for each
probe, and over all unitary interactions.  As opposed to the last
method, which characterizes the nonclassicality of $\hat\rho_0$ and
$\hat\rho_1$ by a {\it single number}, this method characterizes the
same via a complete curve, say,
\begin{equation}
{\cal C}\,=\,\Big\{\,{\cal I}_{\rm opt}(d)\;\Big|\;
0\le d\le1\,\Big\}\;.
\label{Fender}
\end{equation}

Equations~(\ref{Pearl}) and (\ref{Fender}) encapsulate the essence of
what this paper is about.\footnote{Our emphasis of these options for
expressing the information--disturbance principle should not be taken 
to preclude the possibility that {\it weaker\/} bounds along the same
lines might also be useful. For instance, an unachievable bound in
Eq.~(\ref{Pearl}) may also be of interest for various purposes.  For
an application along these lines related to quantum cryptography,
see Ref.~\cite{Luetkenhaus96}.}
Using any of the measures of the preceding
subsections, or using any measures left yet unconsidered, these
equations define quite rigorously what is meant by the phrase
``information gain and quantum state disturbance go hand in hand in
quantum theory.''

In Section~4, we shall carry out the {\it information--disturbance 
principle\/} program in some detail for two pure quantum states.  This
gives something of a flavor for how the program goes in the general 
case, but it also hints at the calculational difficulties that
crop up even in the simplest of cases.  Before pursuing this, however,
we explore in the next section the significance of this principle
within the historical context.  

\section{Historical Context}

In this Section, we shall briefly explore the connection between the
ideas developed here and some common prejudices concerning 
measurements and disturbance.

It is often said that the {\it Heisenberg uncertainty relations\/} are
the ultimate reason that Eve's measurements disturb Alice's 
system in quantum key distribution.\footnote{One notable exception 
is Ref.~\cite{Ekert94}, which asserts, ``\ldots\ quantum cryptography
\ldots\ relies on the impossibility of ascribing definite values to 
noncommuting variables in order to assure secrecy of communication.''}
To the extent that this statement can be given a precise meaning, it 
is rather beside the point in this context.  The Heisenberg relations
concern our conceptual and experimental inability to simultaneously
``get hold'' of two  classical observables, such as a particle's
position $x$ and momentum $p$.  Thus they concern our inability to 
ascribe {\it classical\/} states of motion to quantum mechanical 
systems.  This remains true of any of the standard meanings for the 
phrase ``uncertainty relation''---that is, whether one is speaking of 
a semiclassical argument like Heisenberg's original one 
\cite{Heisenberg27}, the well-known textbook version of the relation 
due to Robertson \cite{Robertson29}, or instead rigorous bounds on the
simultaneous measurability of conjugate variables \cite{Arthurs88}.

Moreover, the Robertson version of the relation has nothing to say
about the idea that observing one classical variable causes a 
``disturbance'' to the other.  In particular, upon following
its derivation carefully, one finds nothing remotely resembling 
the picture of the Heisenberg argument \cite{Peres93b}.  What is found instead is that, when many copies of a system are prepared in the 
same quantum state $\psi(x)$, if one makes enough position 
measurements on the copies to get an estimate of $\Delta x$, and 
similarly makes enough momentum measurements on different(!)\ copies 
to get an estimate of $\Delta p$, then one can be assured that the 
product of those numbers will be no smaller than $\hbar/2$.  No 
reciprocal ``disturbance'' of any kind is involved here, since
$\hat x$ and $\hat p$ are measured on different systems.

The source of the misunderstanding is perhaps typified by statements
such as the following from Wolfgang Pauli, collecting a set of
thoughts now ingrained in the folklore of the theory
\cite[p.~132]{Pauli95}:
\begin{quote}
\baselineskip=11pt
The indivisibility of elementary quantum processes (finiteness of the 
quantum of action) finds expression in an indeterminacy of the 
interaction between instrument of observation (subject) and the system
observed (object), which cannot be got rid of by determinable
corrections.  It is therefore only the experimental arrangement that
defines the physical state of the system, whose characterization thus 
essentially involves some knowledge about the system.  For every act
of observation is an interference, of undeterminable extent, with
the instruments of observation as well as with the system observed,
and interrupts the causal connection between the phenomena preceding 
and succeeding it.  The gain of knowledge by means of an observation 
has as a necessary and natural consequence the loss of some other 
knowledge.  The observer has however the free choice, corresponding
to two mutually exclusive experimental arrangements, of determining
{\it what\/} particular knowledge is gained and what other knowledge
is lost (complementary pairs of opposites).  Therefore every
irrevocable interference by an observation with the sources of
information about a system alters its state, and creates a new 
phenomenon in Bohr's sense.
\end{quote}

These sorts of ideas, folklore or not, have very little to do with the
limits of what can happen to quantum states when information is
gathered about their identity.  The theme of the approach suggested by
quantum cryptography differs from that used in formulating the 
Heisenberg relations in that it makes no reference to conjugate or 
complementary variables.  The only elements entering the 
considerations are related to the quantum states themselves.  Because
of this, one is forced to contemplate a notion of state disturbance 
that is purely quantum mechanical in character, making no reference
to the semi-classical considerations that drove early conceptual work 
on the theory.

Indeed, the main contribution of quantum cryptography to fundamental
physical understanding has been the lesson that that we must focus on 
{\it multiple\/} (non\-orthogonal) quantum states for the 
construction of an information--disturbance principle.\footnote{The
first appearance of this idea seems to be in the work of Stephen 
Wiesner circa 1970.  It was written up in a manuscript titled 
``Conjugate Coding'' submitted to {\it IEEE Transactions on 
Information Theory}, but remained unpublished until much later
\cite{Wiesner83}.}
Once the focus is shifted from conceptually gauging disturbance in
terms of classical variables to gauging it in terms of the the quantum states themselves, there is no necessary disturbance whatsoever
connected to the process of measurement---{\it if one is talking 
about a single quantum state\/} \cite{Kraus83}.  For instance,
suppose Eve encounters only
a single pure state $|\psi\rangle$ in our scenario---rather than one
of a set of two nonorthogonal states---and she makes a measurement
directly on Alice's system corresponding to a set of orthogonal
projectors $\hat\Pi_b=|b\rangle\langle b|$.  Then there is no physical
principle that can prevent her from using a measurement interaction
that returns Alice's quantum system back to the state $|\psi\rangle$
at its completion.  This is a simple consequence of the fact that
Eve knows the identity of the incoming state.  Her general inability
to predict her measurement outcome cannot erase this---it does not 
matter how random the outcome of the measurement turns out to be. 
Nor does it matter that a second measurement of the same observable
may reveal a completely different outcome than the first.

The idea of a necessary disturbance to quantum states upon measurement
only becomes effective when that measurement's ``random'' outcome
reveals some statistical information about the identity of the
unknown quantum state.  Also crucial here is that the set of possible 
states, even though known to Eve, have at least some elements 
nonorthogonal to the others.  This makes a classical representation of the states in terms of mutually existent objective properties 
impossible.  This is the legacy of quantum cryptography.

What does it mean to say that the states are disturbed in and
of themselves without reference to classical variables?  It means
quite literally that Alice faces a loss of predictability about the
outcomes of further measurements on her system whenever an
information gathering eavesdropper has intervened.  Take as an
example the case where $\hat\rho_0$ and $\hat\rho_1$ are nonorthogonal pure states.  Then, for each of these, there exists one nontrivial 
observable for which Alice can predict the outcome with complete
certainty---namely the projectors parallel to $\hat\rho_0$ and 
$\hat\rho_1$.  That is to say, she can ask the system, ``Are you in
the state I prepared you in,'' and it will answer yes with certainty.
However, after her system passes into the black box occupied by 
Eve and returns,  she will no longer be able to predict completely 
the outcomes of both these measurements.  This is the real content of
these ideas: the disturbance is to Alice's description of her system 
and what she can predict of it.

\section{Pure States}

We shall devote ourselves in this Section to working out, for two
nonorthogonal pure states, one of the many tradeoff principles
discussed in Section 2.  The particular one considered is motivated
by its overall simplicity, not by its operational
significance for quantum cryptography.  The overriding concern is
that---in this way---we will be able to carry out the majority of the 
steps analytically.  Consequently, in this exercise, we will be less 
reliant on various numerically motivated conjectures than was the
case for previous endeavors \cite{Fuchs96b}.  It is hoped that this 
procedure will lead to insight into the general structure of more
practically-motivated tradeoff principles, such as that considered in
Ref.~\cite{Fuchs96b}.\footnote{The tradeoff principle there is based
on the same disturbance measure as considered here.  However the
information measure is the actual Shannon information accessible to
Eve.  Under certain circumstances and certain assumptions, these two
quantities are sufficient for defining a privacy amplification
procedure for quantum key distribution \cite{Brassard96T} in the
context of the BB84 protocal \cite{Bennett84}.}

Let us jump into the example.  The two quantum states that Alice 
prepares are $\hat\rho_0=|0\rangle\langle0|$ and
$\hat\rho_1=|1\rangle\langle1|$ where
\begin{eqnarray}
|0\rangle &=& \cos\alpha\,|a_0\rangle\,+\,\sin\alpha\,|a_1\rangle
\label{Hemo}\\
|1\rangle &=& \sin\alpha\,|a_0\rangle\,+\,\cos\alpha\,|a_1\rangle\;,
\label{Globin}
\rule{0mm}{5mm}
\end{eqnarray}
are vectors on a two-dimensional Hilbert space ${\cal H}_A$.
Here $|a_0\rangle$ and $|a_1\rangle$ form an orthonormal basis on
${\cal H}_A$.  Algebraically, the distinction between these states is
parameterized by the single number
\begin{equation}
S=\langle 0|1\rangle=\sin 2\alpha\;.
\label{Nietsche}
\end{equation}

As already discussed in great detail, Eve---in an attempt to gather 
information about the identity of the state Alice prepared---will 
interact Alice's system with a probe described by the states on a 
Hilbert space ${\cal H}_E$.  This will lead to Alice being left
finally in possession of a system described by a density operator
$\hat\rho^{\scriptscriptstyle {\rm A}}_s$.  Similarly Eve will be left
with her probe being described by some density operator
$\hat\rho^{\scriptscriptstyle {\rm E}}_s$.  In each case, the value of
$s$ depends on Alice's particular preparation.

The tradeoff principle we shall consider in this Section is based on
the information-gain and disturbance measures of 
Eqs.~(\ref{LimaPeru}) and (\ref{dingus}), which here reduce to
\begin{equation}
P_e(\hat U)\,=\,\frac{1}{2}\,-\,\frac{1}{4}\,{\rm tr}\Bigl|\,
\hat\rho^{\scriptscriptstyle {\rm E}}_1-
\hat\rho^{\scriptscriptstyle {\rm E}}_0\,\Bigr|
\label{Hubba}
\end{equation}
and
\begin{equation}
D(\hat U)\,=\,1\,-\,\frac{1}{2}\langle 0|
\hat\rho^{\scriptscriptstyle {\rm A}}_0|0\rangle
\,-\,\frac{1}{2}\langle 1|\hat\rho^{\scriptscriptstyle
{\rm A}}_1|1\rangle\;,
\label{Bubba}
\end{equation}
respectively.  That
is to say, the ``information'' measure is taken to be the best 
possible probability of error that Eve can encounter when trying to 
guess the identity of Alice's preparation; the disturbance measure is
taken to be the average probability that Alice will detect a 
discrepancy if she performs a test corresponding to the actual state 
she sent into the black box.

The particular way we shall combine the two ingredients,
Eqs.~(\ref{Hubba}) and (\ref{Bubba}), is in finding a curve as in 
Eq.~(\ref{Fender}).  That is to say, for fixed
\begin{equation}
D(\hat U)=d\;,
\label{BadFinger}
\end{equation}
and
\begin{equation}
P_e^{\rm opt}(d)=\min_{\hat U}P_e(\hat U)\;,
\end{equation}
where the minimization is performed subject to the constraint, we seek
to find the curve
\begin{equation}
{\cal C}\,=\,\Big\{\,P_e^{\rm opt}(d)\;\Big|\;
0\le d\le d_0\,\Big\}\;.
\label{WebSpud}
\end{equation}
The constant $d_0$ denotes the minimal disturbance that need be
incurred in the process of finding the maximal allowed information.
Thus we are assuming that the disturbance is never so large as to be
an unnecessary disturbance.\footnote{This assumption is not necessary;
it is made only for the purpose of relaxing the need to make
repetitive conditional statements later.}

This plan requires that we be able to perform the
constrained variation necessary for it.  We tackle this problem in
several steps.  The first is in finding the simplest general 
description of the probe and its unitary interaction with Alice's 
system.  The second is in calculating all the relevant quantities of
the tradeoff relation for a fixed interaction on Eve's part.  
The third is making a two-state probe plausible by analyzing
a set of variational equations.  Finally the curve is optimized based
on the assumption of a two state probe.

\subsection{The Probe and Interaction}

We set the wheels turning by noticing that there is no utility 
in Eve allowing her probe to start off in a mixed state.  This is 
because any mixed state for Eve's probe can always be thought of as 
arising from a partial trace over the degrees of freedom of a larger 
probe prepared in a pure state \cite{Jozsa94a}.  Since the 
dimensionality of the probe's Hilbert space ${\cal H}_E$ is a free 
variable anyway, this choice will cause no loss in generality.

But then, however, there is no use in declaring the initial state of
the probe in our variations:  any pure state can be transformed 
unitarily into any other pure state.  This forms the reason why we 
have made $\hat U$ the only explicit argument of the various 
information and disturbance measures:  it encodes both the probe's
dimensionality and the particular interaction it undertakes.

Let us now focus on saying something about the dimensionality of
${\cal H}_E$.  The interaction $\hat U$ takes its formal description 
as a unitary operator on ${\cal H}_A\otimes{\cal H}_E$.  Supposing,
as just discussed, the probe starts off in some standard pure state 
$|\psi\rangle$, we can describe the outcome of this interaction as 
follows:
\begin{equation}
|s\rangle|\psi\rangle\;\longrightarrow\;
|\Psi^{\scriptscriptstyle {\rm AE}}_s\rangle=
\sum_{n=0}^1\sqrt{\lambda^s_n}\,|A^s_n\rangle|E^s_n\rangle\;,
\label{NubbleKnees}
\end{equation}
$s=0,1$, and where $|\Psi^{\scriptscriptstyle {\rm AE}}_s\rangle$ is
written in Schmidt polar form \cite{Peres93b} for orthonormal bases 
$|A^s_n\rangle$ and $|E^s_n\rangle$ (parameterized by Alice's 
preparation $s$) on ${\cal H}_A$ and ${\cal H}_E$ respectively.  The
bases and the constants $\lambda^s_n$ will of course depend on the
particular unitary operation $\hat U$ used.  The important thing to
note, however, is that the number of nonzero $\lambda^s_n$ is
determined by the dimensionality of ${\cal H}_A$.

Writing the post-interaction states in Schmidt polar form is
particularly convenient because then one easily sees that the 
state of the system finally left in Eve's possession is of the form
\begin{eqnarray}
\hat\rho^{\scriptscriptstyle {\rm E}}_s
&=&
{\rm tr}_{\scriptscriptstyle {\rm A}}
\Big(|\Psi^{\scriptscriptstyle {\rm AE}}_s\rangle\langle
\Psi^{\scriptscriptstyle {\rm AE}}_s|\Big)
\nonumber\\
&=&
\lambda^s_0|E^s_0\rangle\langle E^s_0|\,+\,\lambda^s_1|E^s_1\rangle
\langle E^s_1|\;.
\rule{0mm}{4mm}
\label{RunkleBean}
\end{eqnarray}
Similarly the states passed back to Alice are of the form
\begin{eqnarray}
\hat\rho^{\scriptscriptstyle {\rm A}}_s
&=&
{\rm tr}_{\scriptscriptstyle {\rm E}}
\Big(|\Psi^{\scriptscriptstyle {\rm AE}}_s\rangle\langle
\Psi^{\scriptscriptstyle {\rm AE}}_s|\Big)
\nonumber\\
&=&
\lambda^s_0|A^s_0\rangle\langle A^s_0|\,+\,\lambda^s_1|A^s_1\rangle
\langle A^s_1|\;.
\rule{0mm}{4mm}
\end{eqnarray}

Eq.~(\ref{RunkleBean}) tells us, rather interestingly, that we never
need consider probes for Eve with Hilbert space dimension greater
than four \cite{Fuchs96b}.  This follows because, at most, all four
of the $|E^s_n\rangle$ can be linearly independent vectors.  This 
means that the relevant states accessible to Eve can never act on more
than the span of these four vectors.  The states
$\hat\rho^{\scriptscriptstyle {\rm E}}_0$ and
$\hat\rho^{\scriptscriptstyle {\rm E}}_1$ are thus always confined to
actions on (effectively) four-dimensional Hilbert spaces at the 
outset.

It was found numerically---but not proven---for the tradeoff relation 
considered in Ref.~\cite{Fuchs96b} that, in the end, a two-state
probe always sufficed for defining the optimal tradeoff curve.  In 
certain ways, this was rather surprising: Eve's probe never need take 
advantage of all the Hilbert space allotted it.  Therefore it would be
very nice to demonstrate the same effect rigorously for the present
tradeoff relation.  Unfortunately, this goal has yet to be found
attainable: the algebra becomes quite unwieldy.  The best that has
been demonstrated (assuming the various mild assumptions to be
made below) is that a two-state probe makes the tradeoff
stationary.  However, this does not demonstrate that a two-state
probe is globally optimal; it only demonstrates that a two-state
probe is locally the best, the worst, or an inflection point.  There
are indeed other local optima, though---based on numerical
simulations---none appear to be the global optimum.

In what follows, we shall sidestep the issue of trying to show that,
{\it starting with a four-dimensional probe\/}, a two-dimensional one
suffices to make the variation stationary.  There appears to be 
nothing to be gained from this exercise other than mathematical
heartache.  Instead, here, we shall start off within the restricted
scenario of allowing ${\cal H}_E$ to be up to {\it three\/}
dimensional.  This allows for a much easier path to the essential 
physics that a two-state probe suffices for the construction of the 
optimal (stationary, to be more precise) $P_e$--$D$ tradeoff curve.  
It also cleans up the mathematics in this
paper considerably and helps make the exposition tractable.  It must
be emphasized, however, that by making this restriction, we cut off 
the possibility of {\it proving\/} that the curve we shall ultimately 
obtain is actually optimal---though, based on numerical work, we
certainly believe it is.  Full rigor or a better methodology for
tackling this problem must remain the subject of future work.

All that said, let us take ${\cal H}_E$ to be a three-dimensional
Hilbert space and let us equip it with an orthonormal basis 
$|e_\beta\rangle$, $\beta=x,y,z$.  (Ultimately we will choose this
basis to be as convenient as possible.)  This gives a
natural basis for the Hilbert space ${\cal H}_A\otimes{\cal H}_E$,
namely the tensor product basis $|a_m\rangle|e_\beta\rangle$, $m=0,1$
and $\beta=x,y,z$.  The set of all possible unitary operators on this
six dimensional Hilbert space is quite large:  without some further
symmetry arguments for reducing its size, our variational problem will
remain quite intractable.  To this end, let us start building some
notation.

Let us write the action of each unitary operator $\hat U$ in our
variational in the following way:
\begin{equation}
\hat U |a_m\rangle|\psi\rangle\,=\,\sum_{n=0}^1|a_n\rangle|R_{mn}
\rangle\;,
\label{Fizzlebo}
\end{equation}
where
\begin{equation}
|R_{mn}\rangle=\sum_\beta U_{mn\beta}|e_\beta\rangle\;.
\label{Nublette}
\end{equation}
Since Eve's probe starts off in the standard state $|\psi\rangle$, we
need not be concerned with the action of $\hat U$ on the remainder of
the Hilbert space.  Note that unitarity requires that
\begin{equation}
\sum_{n,\beta}U_{mn\beta}^*U_{m'n\beta}=\delta_{m,m'}\;.
\label{GeneAutrey}
\end{equation}

At least one thing should be clear at this point.  If Alice were to
relabel her orthonormal basis $|a_m\rangle$ by interchanging $0$
and $1$, this would only lead to a relabeling of her two input states.
The algebraic relation between the input states, Eq.~(\ref{Nietsche}),
remains invariant.  Thus one should expect that Eve will never need
to break that symmetry in her apparatus to ferret out the best
tradeoff.  Therefore, we may fairly safely assume that 
under the interchange $0\leftrightarrow1$, the probe's states 
(relative to the basis states of Alice's system) keep the same
algebraic relations among themselves.  That is to say, we may assume
that the operator $\hat U$ is always such that all inner products 
$\langle R_{m'n'}|R_{mn}\rangle$ remain invariant under an interchange of $0\leftrightarrow1$ in the indices.

This already places severe restrictions on the set of unitary 
operators that we need consider for the variation.  However, we can
still go a few steps further in restriction.  For this let us write
the input states in Eqs.~(\ref{Hemo}) and (\ref{Globin}) as
\begin{equation}
|s\rangle=\sum_{m=0}^1 c_{sm}|a_m\rangle\;.
\end{equation}
Then, we can rewrite Eq.~(\ref{NubbleKnees}) as
\begin{equation}
|\Psi^{\scriptscriptstyle {\rm AE}}_s\rangle=\sum_{n=0}^1
|a_m\rangle|R_n^s\rangle\;,
\end{equation}
where
\begin{equation}
|R_n^s\rangle=\sum_{m=0}^1 c_{sm}|R_{mn}\rangle\;.
\label{SnotCake}
\end{equation}
In this notation, the final states accessible to Eve for 
information gathering measurements are
\begin{equation}
\hat\rho^{\scriptscriptstyle {\rm E}}_s=\sum_{n=0}^1
|R_n^s\rangle\langle R_n^s|\;.
\label{yawn}
\end{equation}

Equation~(\ref{yawn}) is of particular interest because it tells us
that we can freely adjust the phase of all the vectors $|R_n^s\rangle$ with no effect whatsoever on the states received by Eve.  This
translates into the freedom to set the phase of the $|R_{mn}\rangle$
as we wish, as long as we do it to $|R_{00}\rangle$ and 
$|R_{10}\rangle$ in unison and as long as we do it to $|R_{01}\rangle$
and $|R_{11}\rangle$ in unison.  (This follows from the requirement
that Eq.~(\ref{SnotCake}) hold.)

To see where this is going, let us write the four equations implicit
in Eq.~(\ref{Nublette}) as:
\begin{eqnarray}
|R_{00}\rangle &=& X_1 |e_x\rangle + X_2 |e_y\rangle + X_3 |e_z\rangle
\\
|R_{01}\rangle &=& X_4 |e_x\rangle + X_5 |e_y\rangle + X_6 |e_z\rangle
\rule{0mm}{5mm}\\
|R_{10}\rangle &=& X_7 |e_x\rangle + X_8 |e_y\rangle + X_9 |e_z\rangle
\rule{0mm}{5mm}\\
|R_{11}\rangle&=&X_{10}|e_x\rangle+X_{11}|e_y\rangle+X_{12}|e_z\rangle
\;.\rule{0mm}{5mm}
\end{eqnarray}
In terms of these new variables, the unitarity relation
Eq.~(\ref{GeneAutrey}) becomes,
\begin{equation}
\sum_{k=1}^6 X_k^* X_k=\sum_{k=7}^{12} X_k^* X_k=1\;,
\label{HandGum}
\end{equation}
and
\begin{equation}
\sum_{k=1}^6 X_k^* X_{k+6}=0\;.
\label{WristBurn}
\end{equation}

What we would like to do is use the freedom in the arbitrary basis
$|e_\beta\rangle$, the symmetry requirements, and the extra phase
freedom just mentioned to make the $X_k$, $k=1,\ldots,12$ as simple
looking as possible.  To do this, we first choose the phase of
$|R_{11}\rangle$ so that $X_{12}$ is real.  
Next, we recognize the fact that $|R_{01}\rangle$ and $|R_{10}\rangle$ must be of the same length.  This follows because we require
\begin{equation}
\langle R_{01}|R_{01}\rangle=\langle R_{10}|R_{10}\rangle\;.
\end{equation}
Therefore, we may choose the basis states $|e_\beta\rangle$ so that:
$X_4=X_8$, $X_5=X_7$,  $X_4$ and $X_5$ are both real, and
$X_6=X_9=0$.  Finally we use the phase freedom in 
$|R_{00}\rangle$ and $|e_z\rangle$ to make both $X_{2}$ and $X_{3}$
real.  In particular, we take the opportunity to make $X_{3}$ and
$X_{12}$ of the same sign.

Thus the situation is, at present, the following
\begin{eqnarray}
\!\!\!\!\!\!\!|R_{00}\rangle \!\!\!&=&\!\!\! X_1
e^{i\theta_1} |e_x\rangle + X_2
|e_y\rangle + X_3 |e_z\rangle
\\
\!\!\!\!\!\!\!|R_{01}\rangle \!\!\!&=&\!\!\!
X_4 |e_x\rangle + X_5 |e_y\rangle 
\rule{0mm}{5mm}\\
\!\!\!\!\!\!\!|R_{10}\rangle \!\!\!&=&\!\!\!
X_5 |e_x\rangle + X_4 |e_y\rangle 
\rule{0mm}{5mm}\\
\!\!\!\!\!\!\!|R_{11}\rangle \!\!\!&=&\!\!\!
X_{10} e^{i\theta_{10}}|e_x\rangle +
X_{11}e^{i\theta_{11}}|e_y\rangle + X_{12}|e_z\rangle\;.
\rule{0mm}{5mm}
\end{eqnarray}
In these equations and hereafter we treat all the $X_k$ as real, 
making all phases explicit in the factors $e^{i\theta_k}$.

Can we go the remainder of the way and justify setting 
$e^{i\theta_1}$, $e^{i\theta_{10}}$, and $e^{i\theta_{11}}$ all to
be real numbers?  Actually, the answer is yes, though it should
have been quite plausible in the first place: since the input states,
Eqs.~(\ref{Hemo}) and (\ref{Globin}), involve only real coefficients,
there is no reason to expect complex numbers to be necessary for
describing Eve's probe.  Let us prove this now.

The symmetry requirements
\begin{eqnarray}
\langle R_{10}|R_{00}\rangle &=& \langle R_{01}|R_{11}\rangle
\\
\rule{0mm}{4mm}
\langle R_{00}|R_{11}\rangle &=& \langle R_{11}|R_{00}\rangle
\end{eqnarray}
respectively, give
\begin{equation}
X_5 X_1 e^{i\theta_1} + X_4 X_2 =  X_4 X_{10} e^{i\theta_{10}} +
X_5 X_{11}e^{i\theta_{11}}\;,
\label{monkey}
\end{equation}
and that the quantity
\begin{equation}
X_1 X_{10} e^{i(\theta_{10}-\theta_1)} + X_2 X_{11}e^{i\theta_{11}}
\label{mothers}
\end{equation}
is real.  On the other hand, Eq.~(\ref{WristBurn}) requires
\begin{equation}
X_1 X_5 e^{i\theta_1} + X_2 X_4 + X_4 X_{10} e^{i\theta_{10}} +
X_5 X_{11}e^{i\theta_{11}}=0\;.
\label{milk}
\end{equation}
Equations (\ref{monkey}) and (\ref{milk}) taken together imply that
we must have
\begin{equation}
X_1 X_5 e^{i\theta_1} + X_2 X_4 = 0\;.
\end{equation}
Thus $e^{i\theta_1}=\pm1$.  By the same token, we must also have
$e^{i(\theta_{11}-\theta_{10})}=\pm1$.  Finally, these two facts
along with the reality of the quantity in Eq.~(\ref{mothers}) gives
that both $e^{i\theta_{11}}$ and $e^{i\theta_{10}}$ are real.
That is to say, we have just proven that by suitable choice of basis
and phase, all the $X_k$ may be assumed real without loss of
generality.

The four $|R_{mn}\rangle$ can be visualized easily as vectors
in a three-dimensional real vector space.  In fact, by making a
sketch, one immediately sees that the symmetry requirements taken
together uniquely fix the components of the vectors up to a freedom
of relative sign between $X_3$ and $X_{12}$.  However, this particular
freedom was already taken away by fiat above.  Therefore, we set
$X_3=X_{12}$.  Absorbing all the remaining phase freedom back into
the variables, we have in the final accounting:
\begin{eqnarray}
|R_{00}\rangle &=& X_1 |e_x\rangle + X_2 |e_y\rangle + X_3 |e_z\rangle
\\
|R_{01}\rangle &=& X_4 |e_x\rangle + X_5 |e_y\rangle 
\rule{0mm}{5mm}\\
|R_{10}\rangle &=& X_5 |e_x\rangle + X_4 |e_y\rangle 
\rule{0mm}{5mm}\\
|R_{11}\rangle &=& X_2 |e_x\rangle + X_1 |e_y\rangle + X_3 |e_z\rangle
\;.\rule{0mm}{5mm}
\end{eqnarray}

The unitarity relations, Eqs.~(\ref{HandGum}) and (\ref{WristBurn}),
after all these simplifying assumptions, become
\begin{equation}
X_1^2+X_2^2+X_3^2+X_4^2+X_5^2=1\;,
\end{equation}
and
\begin{equation}
X_1 X_5 + X_2 X_4 = 0\;.
\end{equation}
This means that in our variational problem, we need only search over
three free parameters:  this is a substantial savings over approaching
the problem blindly!  We may parameterize the remaining $X_k$ in the
following way:
\begin{eqnarray}
X_1 &=& \cos\lambda\cos\theta\cos\phi\nonumber\\
X_2 &=& \cos\lambda\cos\theta\sin\phi\nonumber\\
X_3 &=& \sin\lambda \label{CherryTree}\\
X_4 &=& \cos\lambda\sin\theta\cos\phi\nonumber\\
X_5 &=& -\cos\lambda\sin\theta\sin\phi\;.\nonumber
\end{eqnarray}

How is a two-state probe---as opposed to a three-state
probe---expressed in these terms?  Notice that if, for whatever 
reason, $\sin\lambda=0$, then the four $|R_{mn}\rangle$ will all be
confined to the same two-dimensional subspace of ${\cal H}_E$.  Thus,
if at the end of the calculation we find that $\lambda$ must be an
integer multiple of $\pi$, we will have demonstrated that Eve's probe 
could have been taken to be a two state system in the first place.

\subsection{After the Interaction}

Let us now fix a particular unitary operator $\hat U$ given by
Eq.~(\ref{CherryTree}) and work out all the operators and quantities 
relevant to both Alice and Eve.

We start by finding the density operators
$\hat\rho^{\scriptscriptstyle {\rm E}}_s$, $s=0,1$, accessible to
Eve.\footnote{We list all the components rather than cut to the
chase of the final expression for $P_e(\hat U)$ to save some work on
the part of anyone else that may wish to construct other
information--disturbance tradeoffs.}  Given all the formalism 
developed, it is really just a question of turning the algebraic 
crank.  After a lot of careful algebra, one finds:
\begin{eqnarray}
(\hat\rho^{\scriptscriptstyle {\rm E}}_0)_{xx}
&=&
\frac{1}{2}\cos^2\!\lambda\Big(1+\cos2\alpha\cos2\phi\Big)
\\
\rule{0mm}{9mm}(\hat\rho^{\scriptscriptstyle {\rm E}}_0)_{xy}
&=&
\frac{1}{4}\cos^2\!\lambda\Big(
(\cos\alpha-\sin\alpha)^2\sin2(\phi-\theta)
\nonumber\\
&&
\mbox{}\hspace{4mm}+(\cos\alpha+\sin\alpha)^2\sin2(\phi+\theta)\Big)
\\
\rule{0mm}{9mm}(\hat\rho^{\scriptscriptstyle {\rm E}}_0)_{xz}
&=&
\frac{1}{2}\sin2\lambda\Big(\sin^2\!\alpha\sin\phi\cos\theta
\nonumber\\
&&
\mbox{}\hspace{4mm}+\cos^2\!\alpha\cos\phi\cos\theta
\nonumber\\
&&
\mbox{}\hspace{4mm}+\cos\alpha\sin\alpha\,(\cos\phi-\sin\phi)
\sin\theta\Big)
\\
\rule{0mm}{8mm}(\hat\rho^{\scriptscriptstyle {\rm E}}_0)_{yy}
&=&
\frac{1}{2}\cos^2\!\lambda\Big(1-\cos2\alpha\cos2\phi\Big)
\\
\rule{0mm}{8mm}(\hat\rho^{\scriptscriptstyle {\rm E}}_0)_{yz}
&=&
\frac{1}{2}\sin2\lambda\Big(\cos^2\!\alpha\sin\phi\cos\theta
\nonumber\\
&&
\mbox{}\hspace{4mm}+\sin^2\!\alpha\cos\phi\cos\theta
\nonumber\\
&&
\mbox{}\hspace{4mm}+\cos\alpha\sin\alpha\,(\cos\phi-\sin\phi)
\sin\theta\Big)
\\
\rule{0mm}{5mm}(\hat\rho^{\scriptscriptstyle {\rm E}}_0)_{zz}
&=&
\sin^2\!\lambda\;.
\end{eqnarray}
Hermiticity determines the remainder of the matrix elements for
$\hat\rho^{\scriptscriptstyle {\rm E}}_0$.  The matrix elements for
$\hat\rho^{\scriptscriptstyle {\rm E}}_1$ are given by exactly the
same expressions except that $\cos\alpha$ and $\sin\alpha$ are
interchanged everywhere.

With all the matrix elements of
$\hat\rho^{\scriptscriptstyle {\rm E}}_0$ and
$\hat\rho^{\scriptscriptstyle {\rm E}}_1$ in hand, it is a
straightforward---though tedious---matter to calculate the eigenvalues
of the operator
\begin{equation}
\hat\Gamma'=
\hat\rho^{\scriptscriptstyle {\rm E}}_1-
\hat\rho^{\scriptscriptstyle {\rm E}}_0
\end{equation}
and to arrive finally at an expression for Eve's best-possible 
probability of error.  Using Eq.~(\ref{Hubba}), this is
\begin{equation}
P_e(\hat U)\,=\,\frac{1}{2}\,-\,\frac{1}{2}\cos2\alpha
\sqrt{G(\hat U)}\;,
\label{Lot}
\end{equation}
where
\begin{equation}
G(\hat U)\,=\,\cos^4\!\lambda\cos^2\!2\phi\,+\,\frac{1}{2}
\sin^2\!2\lambda\Big(1-\sin2\phi\Big)\cos^2\!\theta\;.
\label{Sodom}
\end{equation}
Given the particular strategy of enacting the interaction
$\hat U$, Eqs.~(\ref{Lot}) and (\ref{Sodom}) completely characterize 
the ultimate ``information'' available to Eve about Alice's
preparation (as far as is the concern here).

Let us now turn our attention to describing Alice's situation after
her quantum system re\"emerges from Eve's black box.  To this end
we may write the overall post-interaction state as
\begin{equation}
|\Psi^{\scriptscriptstyle {\rm AE}}_s\rangle=\sum_\beta
|Q_\beta^s\rangle|e_\beta\rangle\;,
\end{equation}
where
\begin{equation}
|Q_\beta^s\rangle=\sum_{m=0}^1\sum_{n=0}^1 c_{sm} U_{mn\beta}
|a_n\rangle\;.
\end{equation}
Then
\begin{equation}
\hat\rho^{\scriptscriptstyle {\rm A}}_s=\sum_\beta
|Q_\beta^s\rangle\langle Q_\beta^s|\;.
\label{pucker}
\end{equation}
Again turning the algebraic crank, one finds:
\begin{eqnarray}
(\hat\rho^{\scriptscriptstyle {\rm A}}_0)_{00}
&=&
\cos^2\!\alpha\Big(\cos^2\!\lambda\cos^2\!\theta+\sin^2\!\lambda\Big)
\nonumber\\
&&
\mbox{}\hspace{4mm}+\sin^2\!\alpha\cos^2\!\lambda\sin^2\!\theta
\\
\rule{0mm}{9mm}(\hat\rho^{\scriptscriptstyle {\rm A}}_0)_{01}
&=&
\cos\alpha\sin\alpha\Big(\sin^2\!\lambda+\cos^2\!\lambda\sin2\phi
\cos2\theta\Big)
\nonumber\\
&&
\mbox{}\hspace{4mm}+\frac{1}{2}\cos^2\!\lambda\cos2\phi\sin2\theta
\\
\rule{0mm}{9mm}(\hat\rho^{\scriptscriptstyle {\rm A}}_0)_{11}
&=&
\sin^2\!\alpha\Big(\cos^2\!\lambda\cos^2\!\theta+\sin^2\!\lambda\Big)
\nonumber\\
&&
\mbox{}\hspace{4mm}+\cos^2\!\alpha\cos^2\!\lambda\sin^2\!\theta\;.
\end{eqnarray}
As before, Hermiticity determines the final matrix element of 
$\hat\rho^{\scriptscriptstyle {\rm A}}_0$, and the matrix elements of
$\hat\rho^{\scriptscriptstyle {\rm A}}_1$ are given by the
expressions above except that $\cos\alpha$ and $\sin\alpha$ are
interchanged everywhere.

Making use of Eq.~(\ref{Bubba}), we finally have an expression for
the disturbance forced upon Alice's states:
\begin{eqnarray}
D(\hat U)\!\!
&=&
\!\!\cos^2\!\lambda\left(\sin^2\!\theta\,-\,\frac{1}{2}S\cos2\phi
\sin2\theta\right.
\phantom{\frac{1}{2}S^2(1-\sin}
\nonumber\\
&&
\phantom{\!\!\cos^2\lambda\left(\sin^2\!\theta\right.}
\,+\,\left.\frac{1}{2}S^2\Big(1-\sin2\phi\Big)\cos\!2\theta\right)\;,
\nonumber\\
\end{eqnarray}
where $S$ is the parameter defined by Eq.~(\ref{Nietsche}).

We are now in a position to make the final step toward generating
the curve of Eq.~(\ref{WebSpud}).

\subsection{The Tradeoff Curve}

We can see from Eq.~(\ref{Lot}) that the problem of finding the
smallest $P_e(\hat U)$ as a function of the constraint 
Eq.~(\ref{BadFinger}) boils down to finding the largest $G(\hat U)$
as function of the same constraint.  Let us turn our attention along
these lines.

In order for $G(\hat U)$ to be maximal, it must be at least
stationary with respect to small variations $\delta\lambda$, 
$\delta\phi$, and $\delta\theta$ in $\lambda$, $\phi$, and $\theta$,
respectively.  Thus we must have
\begin{equation}
G_\lambda\delta\lambda+G_\phi\delta\phi+G_\theta\delta\theta=0\;,
\label{Funk}
\end{equation}
where $G_\lambda$ denotes the partial derivative of $G(\hat U)$ with
respect to $\lambda$, etc.  However, not all of these variations
can be made independently; we must satisfy the constraint that
$D(\hat U)$ remain constant.  Therefore, we must have also:
\begin{equation}
D_\lambda\delta\lambda+D_\phi\delta\phi+D_\theta\delta\theta=0\;.
\label{Wagnel}
\end{equation}

Evaluating all the partial derivatives $G_\lambda$,  $D_\lambda$,
$G_\phi$,  $D_\phi$, etc.,  at $\cos\lambda=0$, we can see
that Eqs.~(\ref{Funk}) and (\ref{Wagnel}) are both automatically
satisfied.  This is no surprise, for if $\cos\lambda=0$, then $D=0$
and $P_e=1/2$: when there is no disturbance whatsoever, there can
be no information gain---this is the result established in
Ref.~\cite{Bennett92a}.

Now suppose $\cos\lambda\ne0$ and $\sin2\phi\ne\pm1$, so that, 
regardless of the value of $\theta$, Eve will be able to gain some
information from her interaction.  Under these conditions, a little
algebra shows that $G_\phi\ne0$.  Thus we may transform our problem
into one of unconstrained variation by taking
\begin{equation}
\delta\phi=-\frac{1}{G_\phi}\Big(G_\lambda\delta\lambda
+G_\theta\delta\theta\Big)\;.
\end{equation}
Eliminating this variable from Eq.~(\ref{Wagnel}), we have that all
remaining stationary points must satisfy
\begin{eqnarray}
D_\lambda G_\phi - D_\phi G_\lambda &=& 0
\label{Muncha}
\\
\rule{0mm}{5mm}
D_\theta G_\phi - D_\phi G_\theta &=& 0\;.
\label{Buncha}
\end{eqnarray}
It is easily verified that both of these equations can always be
satisfied by taking $\sin\lambda=0$.  In particular, when 
$\sin\lambda=0$, Eq.~(\ref{Muncha}) is satisfied automatically.
Equation~(\ref{Buncha}) can be satisfied by further adjusting $\phi$
and $\theta$.

The significance of this result is that a two-state probe for Eve is
sufficient to make the information--disturbance tradeoff
stationary.  As already stated, it would be nice to go further and
show that this set of stationary points also leads to a global
maximum for $G(\hat U)$, but that is a much more difficult problem.
In particular, it is clear that there are still other solutions for
Eqs.~(\ref{Muncha}) and (\ref{Buncha}) with {\it neither\/}
$\cos\lambda=0$ nor $\sin\lambda=0$.

Henceforth, we shall rely on this weak result and numerical work to
take it for granted that an optimal interaction $\hat U$ will have
$\sin\lambda=0$.  With this assumption, the remainder of the problem
of finding Eq.~(\ref{WebSpud}) greatly simplifies.  In particular,
the Eq.~(\ref{Sodom}) reduces to
\begin{equation}
G(\hat U)=\cos^2\!2\phi\;.
\end{equation}
To simplify things even further, let us temporarily think of our 
problem as one of minimizing $D(\hat U)$ subject to the constraint 
that $G(\hat U)$ be fixed.  This problem is equivalent to the one
we set out to solve, and is of use here because we will be able
to invert the results of this to get the desired form.  With
this, we have, for fixed $\phi$, that $D(\hat U)$ is minimized 
when
\cite{Fuchs96b}
\begin{equation}
\tan2\theta=\frac{S\cos2\phi}{1-S^2(1-\sin2\phi)}
\end{equation}
and this results in a minimal value of $D(\hat U)$ given by
\begin{equation}
D_{\rm min}
=\frac{1}{2}-\frac{1}{2}\!\left\{S^2\cos^2\!2\phi+
\Big(1-S^2\left(1-\sin2\phi\right)\Big)^{\!2}\right\}^{\!1/2}.
\label{BoogieShoes}
\end{equation}
Notice again that, as $G\rightarrow0$, $D\rightarrow0$ as it should.
Alternatively, the larger the value of $G$, the more information
that is gained about Alice's system.  When $G=1$, so that Eve's
probability of error is as small as it can be, the smallest associated
disturbance that can be given the states is
\begin{equation}
d_0 = \frac{1}{2} - \frac{1}{2}\sqrt{1-S^2+S^4}\;.
\label{MulchMagnet}
\end{equation}

Equation~(\ref{BoogieShoes}) can be inverted to give the solution
we need.  Using Eq.~(\ref{BadFinger}) and a little algebra, we 
obtain:
\begin{equation}
G_{\rm opt}(d)=\frac{4}{\gamma}
\Big(\sqrt{\gamma d(1-d)}-d(1-d)\,\Big)\;,
\label{EdMcMahon}
\end{equation}
where
\begin{equation}
\gamma\;=\;S^2-S^4\;=\;\frac{1}{4}\sin^2\!4\alpha\;.
\end{equation}
Taking $d_0$ as a standard disturbance, we can finally write the
equation of the long sought after curve:
\begin{equation}
P_e^{\rm opt}(d)=\frac{1}{2} - \frac{1}{2}\cos2\alpha
\sqrt{G_{\rm opt}(d)\,}\;,
\label{Abbott}
\end{equation}
where
\begin{equation}
G_{\rm opt}(d)\;=\;2\sqrt{\frac{d(1-d)}{d_0(1-d_0)}\,}\;-\;
\frac{d(1-d)}{d_0(1-d_0)}\;,
\label{Costello}
\end{equation}
and $d$ is restricted to the range $[0,d_0]$.
This completes the exercise of finding the particular
information--disturbance principle given by
Eqs.~(\ref{BadFinger})--(\ref{WebSpud}) for two pure states.

There are at least two (disappointing) things to notice about
Eqs.~(\ref{Abbott}) and (\ref{Costello}).  The first is the energy
that had to be expended in order to work out a tradeoff relation for
one of---surely---the simplest possible cases.  Given the hoped-for
foundational importance of the principle, this is rather curious.
In contrast, the general Heisenberg uncertainty relation (in the form
given by Robertson \cite{Robertson29}) comes about via a simple
application of the Schwarz inequality.

The second thing to note about Eqs.~(\ref{Abbott}) and 
(\ref{Costello}) is the relative complexity of the curve.  Why is it
not linear?  Or, barring that, why are information and disturbance
not simple reciprocals of one another or some other such simple 
relation?  Despite the seeming (observation based) simplicity of the 
information and disturbance measures chosen for this exercise, it may 
just be the case that there are still better measures to be explored.
Perhaps the proper measures to use are the ones that will give rise
to a simple compact expression for the tradeoff principle.

\section{Mixed States}

The deepest understanding of what the information--dis\-turbance
principle has to say about quantum theory as a whole will necessarily
come from the mixed-state analog of these considerations.  For only
then can a direct comparison be made to classical probability
distributions---the object of study in Shannon information theory.
The results of such a comparison are essential for distilling the
cut between classical and quantum theories \cite{Caves96}.

Unfortunately, in moving to the mixed state version of the information--disturbance principle, the mathematical difficulties 
become even more acute than was the case for pure states.  This 
suggests strongly that new techniques or a better formulation of the 
problem are called for. Nevertheless there is one restricted result 
about information gain versus disturbance for mixed states that
brings out an interesting mystery.  This result is known as the 
``no-broadcasting theorem'' \cite{Barnum96a,Fuchs96a}.

Suppose again that a quantum system, secretly prepared by Alice in
one state from the set ${\cal A}\!=\!\{\hat\rho_0,\hat\rho_1\!\}$, is 
dropped into Eve's black box.  The only difference between the present
scenario and our previous considerations is that now we shall suppose 
the black box built solely for the  purpose of {\it broadcasting\/}
or replicating the quantum state onto two separate quantum systems.
That is to say, a state identical to the original should appear in 
each system when it is considered {\it without regard\/} to the other.
We are willing to allow correlation or quantum entanglement between 
the systems.  Can such a black box be built?  If so, then that will 
certainly provide Eve a way to gain information about the mixed state
without causing a detectable disturbance in the system:  Eve need
simply broadcast the state, give Alice one of the systems, and make
a information-gathering measurement on the other.

The standard ``no-cloning'' theorem of Wootters, Zurek, and Dieks
\cite{Wootters82,Dieks82} prohibits broadcasting when the states in 
${\cal A}$ are pure and nonorthogonal; for the only way to have each 
of the broadcast systems 
described separately by a pure state $|\psi\rangle$ is for their joint
state to be $|\psi\rangle|\psi\rangle$; that is to say, $|\psi\rangle$
must be cloned to be broadcast.  When the states in ${\cal A}$ are 
mixed, however, things are not {\it a priori\/} so clear.

A mixed-state no-cloning theorem is not sufficient to demonstrate 
no-broadcasting, for there are many conceivable ways to broadcast the
mixed states $\hat\rho_s$ without the joint state being in the product
form $\hat\rho_s\otimes\hat\rho_s$, the mixed-state analog of
cloning.  As already stated, the systems are also allowed to be 
correlated or entangled in such a way as to give the right marginal
density operators.  For instance, if the density operators 
$\hat\rho_s$ are diagonal in the same basis and have the spectral 
decomposition
\begin{equation}
\hat\rho_s=\sum_b p_s(b) |b\rangle\langle b|\;,
\end{equation}
a set of potential broadcasting states are the highly correlated
joint states
\begin{equation}
\hat R_s=\sum_b p_s(b)\,|b\rangle|b\rangle\langle b|\langle b|\;,
\end{equation}
which, though not of the product form $\hat\rho_s\otimes\hat\rho_s$, 
reproduces the correct marginal density operators.

The general problem, posed formally, is this.  A quantum system AE is 
composed of two parts, A and E, each having an $N$-dimensional
Hilbert space.  System A is secretly prepared in one state from a set 
${\cal A}\!=\!\{\hat\rho_0,\hat\rho_1\!\}$ of two quantum states.  
System E, slated to receive the unknown state, is in a standard
quantum state $\hat\Sigma$.  The initial state of the composite
system AE is the product state $\hat\rho_s\otimes\hat\Sigma$, where
$s=0$ or 1 specifies which state is to be broadcast.  The question
that is being asked is whether there is any physical 
process $\cal E$ (as described in Section 2.1), consistent with the 
laws of quantum theory, that leads to an evolution of the form
\begin{equation}
\hat\rho_s\otimes\hat\Sigma\;\longrightarrow\;
{\cal E}(\hat\rho_s\otimes\hat\Sigma)=\hat R_s\;,
\end{equation}
where $\hat R_s$ is {\it any\/} state on the $N^2$-dimensional 
Hilbert space AE such that
\begin{equation}
{\rm tr}_{\scriptscriptstyle {\rm A}}(\hat R_s)=\rho_s
\;\;\;\;\;\;\;\;\;\mbox{and}\;\;\;\;\;\;\;\;\;
{\rm tr}_{\scriptscriptstyle {\rm E}}(\hat R_s)=\rho_s\;.
\end{equation}
If there is an ${\cal E}$ that satisfies this for both 
$\hat\rho_0$ and $\hat\rho_1$, then the set ${\cal A}$ can be
{\it broadcast}.  A special case of broadcasting is the ``cloning'' 
evolution specified by
\begin{equation}
{\cal E}_c(\hat\rho_s\otimes\hat\Sigma)=\hat\rho_s\otimes\hat\rho_s\;.
\end{equation}

It turns out that despite Eve's best efforts, a physical evolution
can lead to broadcasting if and only if $\hat\rho_0$ and $\hat\rho_1$ 
commute \cite{Barnum96a}.  In this way the concept of broadcasting
makes a very nice communication theoretic cut between commuting and 
noncommuting density operators; this is another way to express the
cut between classical and quantum state descriptions.\footnote{This
is a distinction that has only appeared before in the Holevo bound to accessible information Eq.~(\ref{DuoTang}): the bound is achieved by 
the accessible information if and only if the signal states commute.}
It also turns out that $\cal A$ is clonable if and only if $\rho_0$ 
and $\rho_1$ are identical or orthogonal, i.e., $\rho_0\rho_1=0$
\cite{Barnum96a}.

This result seems to indicate that a key component in an
information--disturbance principle for mixed states will be the
noncommutivity of the states in $\cal A$, not their nonorthogonality.
(For pure states, the properties of orthogonality and commutivity
coincide.) The enticing mystery that arises from the no-broadcasting theorem is the following.  Study of Eqs.~(\ref{Abbott}) and 
(\ref{Costello}) and their analogs in Ref.~\cite{Fuchs96b} reveals 
that information can be gained about the identity of two {\it pure\/}
quantum states without disturbance if and only if they are orthogonal.
(Also see Refs.~\cite{Bennett92a,Busch96}.)  It follows that there
can be information gain without disturbance if and only if the two 
pure states can be cloned.  That is to say, we have the following
logical relationship for pure states in $\cal A$:
$$
\Big(\mbox{(info gain)}\Longrightarrow\mbox{(disturbance)}\Big)\;
\Longleftrightarrow\;\boldmath{\neg}\mbox{(clonable)}\;.
$$%
Because of this, the no-broadcasting theorem might lead one to expect 
an analogous logical relation to hold for mixed states, with 
broadcasting as the relevant concept.  That is, one might expect
that information can be gained without disturbance if and only if the
two states can be broadcast.  This thought, however, is misguided.  
Life again becomes interestingly more complex when it comes to mixed 
states.

A simple example to consider is that of two density operators
$\hat\rho_0$ and $\hat\rho_1$ for which there is a basis in which
both are block diagonal with identical blocks.  That is to say, for
illustration, each $\hat\rho_s$ has a structure something like:
\begin{equation}
\hat\rho_s = \left(\begin{array}{llllllc}
a^s_1 & a^s_2 &  0    &  0    &  0    &  0    &        \\
a^s_3 & a^s_4 &  0    &  0    &  0    &  0    &        \\
 0    &  0    & b^s_1 & b^s_2 & b^s_3 &  0    &        \\
 0    &  0    & b^s_4 & b^s_5 & b^s_6 &  0    & \cdots \\
 0    &  0    & b^s_7 & b^s_7 & b^s_9 &  0    &        \\
 0    &  0    &  0    &  0    &  0    & c^s_1 &        \\
      &       &       &\vdots &       &       & \ddots 
\end{array}\right)\;.
\end{equation}
Suppose furthermore that the density operators are noncommuting on
each of these blocks.  Thus $\hat\rho_0$ and $\hat\rho_1$ themselves
are noncommuting, and it follows that the set $\cal A$ cannot be 
broadcast.

However, if---depending upon the value of $s$---the trace of
these blocks differ, so that, say, $\hat\rho_0$ is more likely to be
in the ``$a$'' block than $\hat\rho_1$, etc., then it follows 
immediately that {\it there are\/} information gathering measurements
that cause no disturbance to these states.  The measurement POVM 
consists of a set of multidimensional projectors onto each of the 
blocks in the $\hat\rho_s$; this measurement can be performed as a 
quantum nondemolition measurement and yet reveals information about 
the identity because of the differing likelihoods of the outcomes.

An interesting open question is the necessary and sufficient
mathematical criteria required of two density operators to insure that
information gathering measurements necessarily disturb the quantum
states.  Securing this result is surely the first step toward the
formulation of a tradeoff principle for mixed states.

\section{Foundations}

To close the paper, we briefly sketch an observation hinting that 
the information--disturbance tradeoff principle itself has something
to say about the foundations of quantum theory. 

Let us return to the Hilbert space ${\cal H}_A\otimes{\cal H}_E$---the
setting for so much of this paper---but, this time, with the 
conviction to play a new game.  Suppose, for whatever reason, we
believe all the standard quantum mechanical axioms sound and beyond 
question, except one.  Namely, we hold suspect the structure of the 
general group of time evolutions to which this vector space may be
submitted.  Perhaps we are simply skeptics, or have always wanted a 
physical mechanism for wave function collapse in isolated systems, or
have some other religious bent that leads us in this direction; the 
motivation is not all that important.

To be relatively general at the outset, we might suppose that this 
group ${\cal T}$ consists of all maps continuous in time that are 
bijections of ${\cal H}_A\otimes{\cal H}_E$ onto itself, and that it 
contains {\it at least\/} all the unitary operations.  In particular, 
we may not wish to tie down the set of evolutions any further than
this---it might contain other linear maps, nonlinear maps, maps
discontinuous with respect to the topology of the vector space, or
what have you.

As it stands, not much can be said about the group ${\cal T}$.  So let
us now ask what would be required of a mapping $\Phi\in{\cal T}$
if it were to be capable of breaking the principles espoused in this
paper.  One thing that pops to mind is the following definition:
\begin{quote}
A mapping $\Phi$ is said to allow {\it illegal
eavesdropping\/} (i.e., information gain without disturbance) if there
are two nonorthogonal states $|s\rangle$, $s=0,1$, in ${\cal H}_A$ and
a standard state $|\sigma\rangle$ in ${\cal H}_E$ such 
that\footnote{All vectors in this definition are assumed normalized.}
\begin{equation}
\Phi\Big(|s\rangle|\sigma\rangle\Big)=|s\rangle|\sigma_s\rangle\;,
\end{equation}
where
\begin{equation}
0\le|\langle\sigma_0|\sigma_1\rangle|<1\;.
\end{equation}
\end{quote}
Note that if such an illegal eavesdropping map were to exist, 
then---since all the other axioms of quantum mechanics are still 
intact---an appropriate measurement on Eve's system alone (in 
principle well after the interaction) would be able to reveal 
information about the state $|s\rangle$ without disturbing it.  This 
follows because the states $|\sigma_0\rangle$ and $|\sigma_1\rangle$ 
are distinct, and by more than just a phase.  It would not be wise to 
base a quantum cryptographic system on the two states $|0\rangle$ and
$|1\rangle$.

Suppose now, however, having grown accustomed to the benefits of
quantum cryptography, we find the existence of such maps simply too unbearable.  We therefore take it as a principle that time evolutions
corresponding to illegal eavesdropping cannot exist.  The question is,
can we still find time evolutions in $\cal T$ that are more
general than those provided by the unitary group?

This is not a completely trivial question: the exclusion of illegal
eavesdropping maps from $\cal T$ looks to be a rather weak
condition at first sight.  In particular, the formulation of the 
eavesdropping maps given in the definition above concerns product 
states only.  The set of product states on
${\cal H}_A\otimes{\cal H}_E$, however, is an infinitesimally small
part of the total space.  For instance, because of the generality
of $\cal T$, one can easily imagine maps on
${\cal H}_A\otimes{\cal H}_E$ that are perfectly well behaved on 
product states, doing just what we expect---i.e., preserving all
inner products---even though their behavior goes completely awry on 
the set of entangled states.  What is to keep these
{\it nonunitary\/} maps from being left in $\cal T$?

It turns out that the group property of $\cal T$, the fact that it
is assumed to include the unitary group as a subgroup, and Wigner's
famous theorem 
on symmetry operations \cite{Wigner59,Bargman64,Peres93b} are enough
to do just this.  If there is a map in ${\cal T}$ that allows the 
increase or decrease of the modulus of {\it any\/} inner product of 
two states in ${\cal H}_A\otimes{\cal H}_E$---it matters not whether
they are product states or entangled states---then we
can construct another map that will use that effect for illegal
eavesdropping.  Thus, if the information--disturbance tradeoff 
principle is to be upheld, $\cal T$ can only contain inner-product 
modulus preserving maps.  And this, in fact, is the premise for 
Wigner's theorem. This theorem states that, allowing the possible
redefinition of phase, all inner-product modulus preserving
maps must be unitary or anti-unitary.  If the maps are to be 
continuous in time, then they must be unitary.

Now, to make this argument complete,
let us show how to construct an illegal eavesdropping map from an
arbitrary time-continuous nonunitary map $\Psi$ on
${\cal H}_A\otimes{\cal H}_E$.  Since $\Psi$ is not unitary, its
action must decrease the inner product (modulus) of at least two
states.\footnote{If it increases any one inner product, then it
must decrease another.  See Ref.~\cite{Peres90b}.}
Call them $|\phi_0\rangle$ and $|\phi_1\rangle$, and let
\begin{equation}
|\phi^\prime_s\rangle=\Psi(|\phi_s\rangle)\;.
\end{equation}
Then we can start out by picking any two states $|0\rangle$ and 
$|1\rangle$ on ${\cal H}_A$ with
\begin{equation}
\langle0|1\rangle=\langle\phi_0|\phi_1\rangle
\end{equation}
and any two states $|\sigma_0\rangle$ and $|\sigma_1\rangle$ on
${\cal H}_E$ with
\begin{equation}
\langle\sigma_0|\sigma_1\rangle=
\frac{\langle\phi^\prime_0|\phi^\prime_1\rangle}{\langle0|1\rangle}\;.
\end{equation}
Finally picking a standard state $|\sigma\rangle$ on ${\cal H}_E$ and 
letting $\hat U_i$ and $\hat U_f$ be any two unitary operators such 
that
\begin{equation}
\hat U_i|s\rangle|\sigma\rangle=|\phi_s\rangle
\end{equation}
and
\begin{equation}
\hat U_f|\phi^\prime_s\rangle=|s\rangle|\sigma_s\rangle\;,
\end{equation}
we have all the tools we need for the job.  Namely, the map
\begin{equation}
\Phi=U_f\circ\Psi\circ U_i\;,
\end{equation}
that is the composition of all the other maps, is an illegal 
eavesdropping map.   Moreover, this map is in $\cal T$ because of its 
group property.

This simple point demonstrates that there are things to be learned
about quantum theory itself by observing how it can be used for
communication and computation.  The particular result here is not 
all that satisfactory in that almost all the structure of quantum
mechanics was taken for granted at the outset.  Moreover, we had to 
require that ${\cal T}$ contain at least the unitary maps before we
could make any progress.\footnote{N.~Gisin has pointed out that
this particular assumption can be done away with by relying on the
strengthened version of Wigner's theorem reported in
Ref.~\cite{Gisin93}.}  Nevertheless, it does provide food for thought
about the directions to which quantum information theory can turn.

\section{Appendix: Error Probability}

In this Appendix, the optimal probability of error
Eq.~(\ref{LimaPeru}) is derived in a previously unpublished way.
Previous derivations \cite{Helstrom67,Helstrom76} have always relied
on the fact that a decision problem was of the main concern; therefore
the derivations have always taken the liberty of assuming that an 
optimal POVM for error probability need only have two outcomes.

We start with an alternative way of writing the classical probability
of error, Eq.~(\ref{Oswald}) \cite{Toussaint72c}, namely,
\begin{equation}
P_e(p_0,p_1)=\frac{1}{2}-K(p_0,p_1)\;,
\end{equation}
where
\begin{equation}
K(p_0,p_1)=\frac{1}{2}\sum_b\Big|\pi_1\,p_1(b)-\pi_0\,p_0(b)\Big|
\end{equation}
is the {\it Kolmogorov variational distance\/} \cite{Kailath67}.
Therefore, finding the optimal value of $P_e(p_0,p_1)$ over all
measurements is equivalent to finding the maximal value of the
Kolmogorov variational distance.

Using the explicit quantum mechanical forms for the probabilities of 
the outcomes the measurement $\{\hat E_b\}$ the Kolmogorov distance
becomes
\begin{eqnarray}
K(p_0,p_1)
&=&
\frac{1}{2}
\sum_b\Big|\pi_1{\rm tr}(\hat\rho_1^{\scriptscriptstyle{\rm E}}
\hat E_b)-\pi_0{\rm tr}(\hat\rho_0^{\scriptscriptstyle{\rm E}}
\hat E_b)\Big|
\nonumber\\
&=&
\frac{1}{2}
\sum_b\Big|{\rm tr}(\hat\Gamma\hat E_b)\Big|\;.
\end{eqnarray}
This follows from the linearity of the trace and the definition of
Eq.~(\ref{BowTie}).  Now let $|\gamma_i\rangle$ denote an orthonormal
eigenbasis for $\hat\Gamma$ and let $\gamma_i$ denote the associated
eigenvalues.  Taking into account Eq.~(\ref{KeyLargo}), we have
\begin{eqnarray}
K(p_0,p_1)
&=&
\frac{1}{2}
\sum_b\left|\sum_{i=1}^N \gamma_i\,\langle\gamma_i|\hat E_b
|\gamma_i\rangle\right|
\nonumber\\
&\le&
\frac{1}{2}
\sum_b\sum_{i=1}^N |\gamma_i|\,\langle\gamma_i|\hat E_b
|\gamma_i\rangle
\nonumber\\
&=&
\frac{1}{2}
\sum_{i=1}^N |\gamma_i|\,\langle\gamma_i|\!\left(\sum_b\hat E_b
\right)\!|\gamma_i\rangle
\nonumber\\
&=&
\frac{1}{2}
\sum_{i=1}^N |\gamma_i|\;.
\end{eqnarray}
The last expression in this, however, is equal by definition to 
${\rm tr}|\hat\Gamma|$.  Thus we have, for all POVMs $\{\hat E_b\}$,
\begin{equation}
K(p_0,p_1)\le\frac{1}{2}{\rm tr}|\hat\Gamma|\;.
\end{equation}
To see that this upper bound can be achieved by some POVM, simply
take the $\hat E_b$ to be the projectors onto an orthonormal 
eigenbasis of $\hat\Gamma$.

\section{Acknowledgments}
I thank several colleagues for enlightening discussions on the ideas
in this paper:
H. Barnum,
M. Boyer,
G. Brassard,
N. Gisin,
L. Levitin,
M.~A. Nielsen,
A. Peres,
J. Preskill,
B. Schumacher,
J. van de Graaf, and
W.~K. Wootters.
In particular, I thank G.~L. Comer for reminding me how to take a
derivative.


\end{document}